\documentstyle[12pt]{article}
\renewcommand{\baselinestretch}{1.5}
\begin{document}
\vskip 72pt
\centerline{\large CAN DILEPTON SUPPRESSION BE SEEN IN HEAVY ION 
EXPERIMENTS ?}
\vskip 36pt
\centerline {\bf {Abhijit Bhattacharyya$^a$\footnote{phys@boseinst.ernet.in}, 
Jan-e Alam$^b$\footnote{jane@veccal.ernet.in}, Sibaji 
Raha$^a$\footnote{sibaji@boseinst.ernet.in} and Bikash 
Sinha$^{b,c}$\footnote{bikash@veccal.ernet.in}}}
\vskip 1pt
\centerline{a) Department of Physics, Bose Institute}
\centerline{93/1 A.P.C.Road}
\centerline{Calcutta 700 009, INDIA}

\centerline{b) Variable Energy Cyclotron Centre}
\centerline{1/AF, Bidhannagar, Calcutta - 700 064, INDIA}

\centerline{c) Saha Institute of Nuclear Physics}
\centerline{1/AF, Bidhannagar, Calcutta - 700 064, INDIA}
\vskip 36pt
\vskip 15 pt

\centerline{\bf {ABSTRACT}}

\noindent
    Effects of finite temperature correction to the
dilepton emission rate in ultrarelativistic heavy ion collisions have been 
looked into. It has been seen that although the $\rho$-peak in the dilepton 
spectra from the hadronic sector is suppressed at very high temperatures, 
almost no effect can be observed in the space time integrated count rate.

\newpage
Very energetic collisions of heavy ions has become an intense field of study 
in recent times. The primary motivation for these investigations is to look 
for a putative new state of matter - Quark Gluon Plasma (QGP). But even more 
generally, the formation of highly excited hadronic matter at high 
temperature/density is an interesting issue in its own right. Dileptons and 
photons emanating from the reaction volume constitute an ideal means for 
studying the property of such hot matter - hadronic or QGP \cite{a,b,c}.

Recently Lee {\it{et.al.}} \cite{d,xx} have calculated the finite temperature 
effects 
on the dilepton spectrum from the hadronic sector. They have used the results 
of QCD sum rule, at finite temperature, to arrive at the conclusion that 
the $\rho$-peak in the dilepton emission rate is suppressed in the hot 
hadronic matter. They have also claimed that this suppression may be observed 
in the experiments. Their result however refers to the static rate {\it{i.e.}} 
number of dileptons coming out from a particular space time point 
(characterised by a single temperature) of the hadronic part of the plasma. 
Unfortunately, there is no experimental way to measure this quantity. In 
experiments, one measures the total number of dileptons coming out of the  
plasma during the entire evolution time (i.e. starting from the 
time of formation upto freezeout). In this brief report we will look into the finite 
temperature effects on the experimental dilepton spectra using the same 
parameterisation as of Lee {\it{et.al.}} \cite{d}.

The most significant channel for the production of dileptons in the QGP sector 
is by annihilation of quark-antiquark pairs ($q{\bar q } \rightarrow \mu^+ 
\mu^-$) and by decays and binary reactions in the hadronic sector 
\cite{b,c,e,f,g,h}. As the emission rates from only the hadronic sector are 
modified by the temperature dependence of the form factors, we will consider 
only this sector in this work, as in \cite{d}. Also, we confine our attention 
to the pion channel alone so as to compare our results with those of ref.
\cite{d}.

The elementary crosssection for pion  annihilation $(\pi^+ \pi^- \rightarrow 
\mu^+ \mu^- )$, assuming vector dominance, is given by \cite{b}
\begin{equation}
{\sigma_\pi}(M) = {F_\pi}^2 (M) \left[ 1 - {{4 {m_\pi}^2} \over M^2}\right]
\sigma (M)
\end{equation}
where

\begin{equation}
\sigma (M) = { {4 \pi} \over 3} {{\alpha^2} \over {M^2}} \left[ 1 + {{2 {m_l}^2}
 \over M^2}\right]  \left[ 1 - {{4 {m_l}^2} \over M^2} \right]^{1/2}
\end{equation}
and $F_\pi$ (M), the pion form factor, is given by

\begin{equation}
{F_\pi}^2 (M) = {m_\rho^4 \over {{(M^2 - m_\rho^2)}^2} + m_\rho^2 
\Gamma_\rho^2}
\end{equation}
where $M$ is the invariant mass, $m_l$  the lepton mass,  $m_\pi$  
 the  pion mass, $m_\rho$  the rho mass and $\Gamma_\rho$  the rho decay width. 

The thermal rate at fixed temperature T can be written as \cite{b,g}

\begin{equation}
{dN \over {d^4 x dM^2 d^2p_T dy}} =  {1 \over {4(2\pi)^5}} M^2 
\left[ 1 - {{4 {m_l}^2} \over M^2} \right] \sigma_\pi(M) e^{-E/T}           
\end{equation}
where $p_T$ is the transverse momentum and $y$ is the rapidity.

We refer to the above expression as the "static rate", implying that the 
spatial/temporal variation of $T$ is not taken into account.

Equation (4) gets modified at finite temperature through the temperature 
dependence of the form factor which, in turn, comes from the temperature 
dependence of the hadron masses, decay widths and coupling constants.      

According to Lee {\it{et.al.}}, \cite {d}, the above quantity gets modified at 
high temperature by a factor of $(1-\delta)$ where $\delta = T^2/6{f_\pi ^2}
$, $f_\pi = 93 MeV$ is the pion decay constant. As a result the height of the 
$\rho$-peak is suppressed. 

Let us now look at the total emission rate of dileptons which is obtained by 
integrating eqn.(4) over the whole space-time volume. So,
\begin{equation}
{dN \over {dM^2 dy}} = \int d^2 p_T d^4 x (1-\delta) {1 \over {4 (2\pi)^5}} 
M^2 \left[ 1 - {{4 {m_l}^2} \over M^2} \right] \sigma_\pi(M) e^{-E/T}           
\end{equation}
In the framework of Bjorken's hydrodynamics \cite{i} the above expression 
takes the form \cite{c}

\begin{equation}
{dN \over {dM^2 dy}} = {\alpha^2 \over \pi^2} \int \tau d\tau r dr  (1-\delta)
M T K_1 (M/T) \left [G(M^2) \Theta (H) + G(M^2) (1 - f_Q) \Theta (M) \right]
\end{equation}
where $f_Q(\tau) =  ({{\epsilon_Q - \epsilon} \over {\epsilon_Q - 
\epsilon_H}})$ is the volume fraction of the quark matter in the mixed 
phase so that $(1-f_Q(\tau))$ is the amount of hadronic matter in the mixed 
phase at a time $\tau$. ${G(M)}^2 = { 1 \over 12} {F_\pi}^2 (M) \left[ 1 - 
{{4 {m_l}^2} \over M^2} \right]$. $\Theta(H)$ and $\Theta(M)$ are the usual 
$\theta$ functions corresponding to the purely hadronic and the mixed phases, 
respectively. If $s_Q$ denotes the entropy density in the QGP phase at 
$T=T_c$, $s_H$ the entropy density in the hadronic phase at $T=T_c$ and $s$ 
the instantaneous entropy density, then $\Theta (H) = \Theta (s_H - s)$ and 
$\Theta (M) = \Theta (s - s_H) \Theta (s_Q - s)$. All other terms have their 
usual meaning. The above equation is the space time integrated rate, the only 
quantity that is accessible in the experiment. 

The freeze-out temperature $T_F$ can not be {\it{a priori}} fixed without 
additional considerations. It is however expected that it should go to lower 
values with increasing mass numbers of the colliding nuclei. For our present 
purpose, we treat it as a parameter and use $T_F=50 MeV$. We have however 
verified that our conclusions (to follow) are quite insensitive to the actual 
values of $T_F$ as long as $T_F \le 140 MeV$.

To summarise, using the parameterization of Lee et.al., we have calculated 
both the static and the space-time integrated emission rate for dileptons 
from the hadronic sector. The static rate has been calculated at $T = 160$ 
and $200 MeV$. It is true that the QCD sum rule, which is based on Operator 
Product Expansion (OPE), is valid only upto the limit where $\delta << 1$. 
But, to see the maximum suppression we have extended it to $\delta \sim 1$ 
({\it{i.e.}} $T = 200 MeV$). In figure ~1a and figure ~1b we have plotted the 
static rate and the space time integrated rate has been plotted in figure~2a 
and figure~2b. As seen from fig.1a and 1b, the $\rho$-peak is suppressed at 
finite temperature. This is the result obtained by Lee {\it {et.al.}} whence 
they have concluded that this suppression may be observed in the experiment. 
However, as we already mentioned in the above the experimentally accessible 
quantity is the total space-time integrated rate (equation 6). We have plotted 
the results of the finite temperature effects on the total emission rates in 
figures 2a and 2b. It can be seen from these two figures that the suppression 
of the $\rho$-peak in the dilepton spectra is almost washed out. The magnitude 
of the surviving difference is so small that one can not expect to observe it 
in the experiment. During the evolution in the hadronic phase, the system 
spends a longer time in the lower temperature configuration where even in the 
static rate the difference is negligible. Most of the differencee in the 
static rate is expected in the mixed phase which is long lived and also at the 
highest temperature achievable in the hadronic phase. But, the amount of 
hadronic matter in the mixed phase is not constant throughout the lifetime of 
the mixed phase, it starts with a very small amount and gradually whole system 
converts to the hadronic matter. Thus the difference in the static rate gets 
softened even in the mixed phase.  

Recently, some calculations have been done in the same  direction \cite{n,o}. 
In ref.\cite{o}, a suppression in the static dilepton emission rate has been 
obtained from a parameterization which assumes that the hadronic modes do not 
propagate beyond $T_c$. In such a situation, the static emission rate at $T_c$ 
is thought to go to zero. This scenario however is not entirely consistent 
with the recent Lattice data \cite{p} and the work of Gocksh \cite{q}, where 
the hadronic modes are argued to survive much beyond $T_c$. But more 
importantly, even for such a strong temperature dependence the difference 
almost gets washed out after space-time integration. 

We thus conclude that in relativistic heavy ion collisions, one can not 
reasonably expect to observe a suppression in the dilepton emission rate due 
to the finite temperature modification of hadronic form factors. The space-time 
integration plays a crucial role in this game. It softens even very strong 
difference which occurs in the static rate. The importance of our work can be 
gauged from the large number of papers that have appeared in the literature in 
the recent times \cite{d,xx,n,o}. All these authors have looked at 
only the static rate and gone on to draw far-reaching conclusions. It has even 
been suggested that the suppression in the dilepton emission rate be used as a 
signature for the partial restoration of chiral symmetry in hot hadronic 
matter formed in ultrarelativistic nuclear collisions \cite{xx}. We have shown 
here that upon a consistent handling of the space-time evolution of the 
system, such effects are almost impossible to discern in actual experiments. 
Our work should thus serve as a caution to the community that neglect of 
space-time integration may lead to optimistic expectations which are far from 
reality. Though the results obtained in this paper are confined to the pion 
annihilation process the main conclusion of this paper that the space-time 
integration softens the difference in the static rate will be valid for all 
the processes. The calculations carried out in this direction, so far, are of 
course based on some model or parameterization. But, keeping in mind the fact 
that the space-time integration kills even very strong difference, one expects 
that the conclusion of this work will not change whatever model one uses.

The work of Abhijit Bhattacharyya was supported, in part, by Department of 
Atomic Energy (Government of India) through K.S.Krishnan fellowship.

\newpage
{\bf{ FIGURE CAPTIONS :}}
\vskip 0.1in

\noindent
1a) Static rate (at constant temperature) of dilepton emission, for $\pi^+ \pi^
- \rightarrow \mu^+ \mu^-$, from the hadronic phase at RHIC energy at a 
temperature $T = 160MeV$. The solid line is for no finite temperature  
correction and the diamonds are with finite temperature 
correction.               

\noindent
1b) Same as 1a but at a temperature $T = 200 MeV$. 

\noindent
2a) Thermal spectra of dileptons, for $\pi^+ \pi^- \rightarrow \mu^+ \mu^-$, 
without transverse expansion from a pion gas at RHIC energy for $T_c = 160 MeV$.
The solid line is for no finite temperature correction and the diamonds are 
with finite temperature correction.               
\noindent
2b) Same as 2a, but for $T_c = 200 MeV$.

\newpage
\begin{figure}
\setlength{\unitlength}{0.240900pt}
\ifx\plotpoint\undefined\newsavebox{\plotpoint}\fi
\sbox{\plotpoint}{\rule[-0.500pt]{1.000pt}{1.000pt}}%
\begin{picture}(1500,1049)(0,0)
\font\gnuplot=cmr10 at 10pt
\gnuplot
\sbox{\plotpoint}{\rule[-0.500pt]{1.000pt}{1.000pt}}%
\put(176.0,68.0){\rule[-0.500pt]{4.818pt}{1.000pt}}
\put(154,68){\makebox(0,0)[r]{$1e-11$}}
\put(1416.0,68.0){\rule[-0.500pt]{4.818pt}{1.000pt}}
\put(176.0,140.0){\rule[-0.500pt]{2.409pt}{1.000pt}}
\put(1426.0,140.0){\rule[-0.500pt]{2.409pt}{1.000pt}}
\put(176.0,182.0){\rule[-0.500pt]{2.409pt}{1.000pt}}
\put(1426.0,182.0){\rule[-0.500pt]{2.409pt}{1.000pt}}
\put(176.0,212.0){\rule[-0.500pt]{2.409pt}{1.000pt}}
\put(1426.0,212.0){\rule[-0.500pt]{2.409pt}{1.000pt}}
\put(176.0,235.0){\rule[-0.500pt]{2.409pt}{1.000pt}}
\put(1426.0,235.0){\rule[-0.500pt]{2.409pt}{1.000pt}}
\put(176.0,254.0){\rule[-0.500pt]{2.409pt}{1.000pt}}
\put(1426.0,254.0){\rule[-0.500pt]{2.409pt}{1.000pt}}
\put(176.0,270.0){\rule[-0.500pt]{2.409pt}{1.000pt}}
\put(1426.0,270.0){\rule[-0.500pt]{2.409pt}{1.000pt}}
\put(176.0,284.0){\rule[-0.500pt]{2.409pt}{1.000pt}}
\put(1426.0,284.0){\rule[-0.500pt]{2.409pt}{1.000pt}}
\put(176.0,297.0){\rule[-0.500pt]{2.409pt}{1.000pt}}
\put(1426.0,297.0){\rule[-0.500pt]{2.409pt}{1.000pt}}
\put(176.0,307.0){\rule[-0.500pt]{4.818pt}{1.000pt}}
\put(154,307){\makebox(0,0)[r]{$1e-10$}}
\put(1416.0,307.0){\rule[-0.500pt]{4.818pt}{1.000pt}}
\put(176.0,380.0){\rule[-0.500pt]{2.409pt}{1.000pt}}
\put(1426.0,380.0){\rule[-0.500pt]{2.409pt}{1.000pt}}
\put(176.0,422.0){\rule[-0.500pt]{2.409pt}{1.000pt}}
\put(1426.0,422.0){\rule[-0.500pt]{2.409pt}{1.000pt}}
\put(176.0,452.0){\rule[-0.500pt]{2.409pt}{1.000pt}}
\put(1426.0,452.0){\rule[-0.500pt]{2.409pt}{1.000pt}}
\put(176.0,475.0){\rule[-0.500pt]{2.409pt}{1.000pt}}
\put(1426.0,475.0){\rule[-0.500pt]{2.409pt}{1.000pt}}
\put(176.0,494.0){\rule[-0.500pt]{2.409pt}{1.000pt}}
\put(1426.0,494.0){\rule[-0.500pt]{2.409pt}{1.000pt}}
\put(176.0,510.0){\rule[-0.500pt]{2.409pt}{1.000pt}}
\put(1426.0,510.0){\rule[-0.500pt]{2.409pt}{1.000pt}}
\put(176.0,524.0){\rule[-0.500pt]{2.409pt}{1.000pt}}
\put(1426.0,524.0){\rule[-0.500pt]{2.409pt}{1.000pt}}
\put(176.0,536.0){\rule[-0.500pt]{2.409pt}{1.000pt}}
\put(1426.0,536.0){\rule[-0.500pt]{2.409pt}{1.000pt}}
\put(176.0,547.0){\rule[-0.500pt]{4.818pt}{1.000pt}}
\put(154,547){\makebox(0,0)[r]{$1e-09$}}
\put(1416.0,547.0){\rule[-0.500pt]{4.818pt}{1.000pt}}
\put(176.0,619.0){\rule[-0.500pt]{2.409pt}{1.000pt}}
\put(1426.0,619.0){\rule[-0.500pt]{2.409pt}{1.000pt}}
\put(176.0,661.0){\rule[-0.500pt]{2.409pt}{1.000pt}}
\put(1426.0,661.0){\rule[-0.500pt]{2.409pt}{1.000pt}}
\put(176.0,691.0){\rule[-0.500pt]{2.409pt}{1.000pt}}
\put(1426.0,691.0){\rule[-0.500pt]{2.409pt}{1.000pt}}
\put(176.0,714.0){\rule[-0.500pt]{2.409pt}{1.000pt}}
\put(1426.0,714.0){\rule[-0.500pt]{2.409pt}{1.000pt}}
\put(176.0,733.0){\rule[-0.500pt]{2.409pt}{1.000pt}}
\put(1426.0,733.0){\rule[-0.500pt]{2.409pt}{1.000pt}}
\put(176.0,749.0){\rule[-0.500pt]{2.409pt}{1.000pt}}
\put(1426.0,749.0){\rule[-0.500pt]{2.409pt}{1.000pt}}
\put(176.0,763.0){\rule[-0.500pt]{2.409pt}{1.000pt}}
\put(1426.0,763.0){\rule[-0.500pt]{2.409pt}{1.000pt}}
\put(176.0,776.0){\rule[-0.500pt]{2.409pt}{1.000pt}}
\put(1426.0,776.0){\rule[-0.500pt]{2.409pt}{1.000pt}}
\put(176.0,786.0){\rule[-0.500pt]{4.818pt}{1.000pt}}
\put(154,786){\makebox(0,0)[r]{$1e-08$}}
\put(1416.0,786.0){\rule[-0.500pt]{4.818pt}{1.000pt}}
\put(176.0,859.0){\rule[-0.500pt]{2.409pt}{1.000pt}}
\put(1426.0,859.0){\rule[-0.500pt]{2.409pt}{1.000pt}}
\put(176.0,901.0){\rule[-0.500pt]{2.409pt}{1.000pt}}
\put(1426.0,901.0){\rule[-0.500pt]{2.409pt}{1.000pt}}
\put(176.0,931.0){\rule[-0.500pt]{2.409pt}{1.000pt}}
\put(1426.0,931.0){\rule[-0.500pt]{2.409pt}{1.000pt}}
\put(176.0,954.0){\rule[-0.500pt]{2.409pt}{1.000pt}}
\put(1426.0,954.0){\rule[-0.500pt]{2.409pt}{1.000pt}}
\put(176.0,973.0){\rule[-0.500pt]{2.409pt}{1.000pt}}
\put(1426.0,973.0){\rule[-0.500pt]{2.409pt}{1.000pt}}
\put(176.0,989.0){\rule[-0.500pt]{2.409pt}{1.000pt}}
\put(1426.0,989.0){\rule[-0.500pt]{2.409pt}{1.000pt}}
\put(176.0,1003.0){\rule[-0.500pt]{2.409pt}{1.000pt}}
\put(1426.0,1003.0){\rule[-0.500pt]{2.409pt}{1.000pt}}
\put(176.0,1015.0){\rule[-0.500pt]{2.409pt}{1.000pt}}
\put(1426.0,1015.0){\rule[-0.500pt]{2.409pt}{1.000pt}}
\put(176.0,1026.0){\rule[-0.500pt]{4.818pt}{1.000pt}}
\put(154,1026){\makebox(0,0)[r]{$1e-07$}}
\put(1416.0,1026.0){\rule[-0.500pt]{4.818pt}{1.000pt}}
\put(176.0,68.0){\rule[-0.500pt]{1.000pt}{4.818pt}}
\put(176,23){\makebox(0,0){$0.5$}}
\put(176.0,1006.0){\rule[-0.500pt]{1.000pt}{4.818pt}}
\put(302.0,68.0){\rule[-0.500pt]{1.000pt}{4.818pt}}
\put(302,23){\makebox(0,0){$0.55$}}
\put(302.0,1006.0){\rule[-0.500pt]{1.000pt}{4.818pt}}
\put(428.0,68.0){\rule[-0.500pt]{1.000pt}{4.818pt}}
\put(428,23){\makebox(0,0){$0.6$}}
\put(428.0,1006.0){\rule[-0.500pt]{1.000pt}{4.818pt}}
\put(554.0,68.0){\rule[-0.500pt]{1.000pt}{4.818pt}}
\put(554,23){\makebox(0,0){$0.65$}}
\put(554.0,1006.0){\rule[-0.500pt]{1.000pt}{4.818pt}}
\put(680.0,68.0){\rule[-0.500pt]{1.000pt}{4.818pt}}
\put(680,23){\makebox(0,0){$0.7$}}
\put(680.0,1006.0){\rule[-0.500pt]{1.000pt}{4.818pt}}
\put(806.0,68.0){\rule[-0.500pt]{1.000pt}{4.818pt}}
\put(806,23){\makebox(0,0){$0.75$}}
\put(806.0,1006.0){\rule[-0.500pt]{1.000pt}{4.818pt}}
\put(932.0,68.0){\rule[-0.500pt]{1.000pt}{4.818pt}}
\put(932,23){\makebox(0,0){$0.8$}}
\put(932.0,1006.0){\rule[-0.500pt]{1.000pt}{4.818pt}}
\put(1058.0,68.0){\rule[-0.500pt]{1.000pt}{4.818pt}}
\put(1058,23){\makebox(0,0){$0.85$}}
\put(1058.0,1006.0){\rule[-0.500pt]{1.000pt}{4.818pt}}
\put(1184.0,68.0){\rule[-0.500pt]{1.000pt}{4.818pt}}
\put(1184,23){\makebox(0,0){$0.9$}}
\put(1184.0,1006.0){\rule[-0.500pt]{1.000pt}{4.818pt}}
\put(1310.0,68.0){\rule[-0.500pt]{1.000pt}{4.818pt}}
\put(1310,23){\makebox(0,0){$0.95$}}
\put(1310.0,1006.0){\rule[-0.500pt]{1.000pt}{4.818pt}}
\put(1436.0,68.0){\rule[-0.500pt]{1.000pt}{4.818pt}}
\put(1436,23){\makebox(0,0){$1$}}
\put(1436.0,1006.0){\rule[-0.500pt]{1.000pt}{4.818pt}}
\put(176.0,68.0){\rule[-0.500pt]{303.534pt}{1.000pt}}
\put(1436.0,68.0){\rule[-0.500pt]{1.000pt}{230.782pt}}
\put(176.0,1026.0){\rule[-0.500pt]{303.534pt}{1.000pt}}
\put(882,-110){\makebox(0,0)[r]{$M(GeV)$}}
\put(882,-350){\makebox(0,0)[r]{{\bf{Figure 1a}}}}
\put(-75,547){\makebox(0,0)[r]{$({dR \over {dM^2d^2p_Tdy}})$}}
\put(882,254){\makebox(0,0)[r]{$T=160MeV$}}
\put(176.0,68.0){\rule[-0.500pt]{1.000pt}{230.782pt}}
\sbox{\plotpoint}{\rule[-0.300pt]{0.600pt}{0.600pt}}%
\put(176,638){\usebox{\plotpoint}}
\multiput(176.00,638.99)(8.615,0.501){11}{\rule{9.600pt}{0.121pt}}
\multiput(176.00,636.75)(106.075,8.000){2}{\rule{4.800pt}{0.600pt}}
\multiput(302.00,647.00)(4.352,0.500){25}{\rule{5.190pt}{0.121pt}}
\multiput(302.00,644.75)(115.228,15.000){2}{\rule{2.595pt}{0.600pt}}
\multiput(428.00,662.00)(2.924,0.500){39}{\rule{3.586pt}{0.121pt}}
\multiput(428.00,659.75)(118.556,22.000){2}{\rule{1.793pt}{0.600pt}}
\multiput(554.00,684.00)(2.058,0.500){57}{\rule{2.589pt}{0.120pt}}
\multiput(554.00,681.75)(120.627,31.000){2}{\rule{1.294pt}{0.600pt}}
\multiput(680.00,715.00)(2.674,0.500){43}{\rule{3.300pt}{0.121pt}}
\multiput(680.00,712.75)(119.151,24.000){2}{\rule{1.650pt}{0.600pt}}
\multiput(806.00,736.50)(1.719,-0.500){69}{\rule{2.193pt}{0.120pt}}
\multiput(806.00,736.75)(121.448,-37.000){2}{\rule{1.097pt}{0.600pt}}
\multiput(932.00,699.50)(0.636,-0.500){193}{\rule{0.914pt}{0.120pt}}
\multiput(932.00,699.75)(124.104,-99.000){2}{\rule{0.457pt}{0.600pt}}
\multiput(1058.00,600.50)(0.624,-0.500){197}{\rule{0.899pt}{0.120pt}}
\multiput(1058.00,600.75)(124.135,-101.000){2}{\rule{0.449pt}{0.600pt}}
\multiput(1184.00,499.50)(0.671,-0.500){183}{\rule{0.954pt}{0.120pt}}
\multiput(1184.00,499.75)(124.019,-94.000){2}{\rule{0.477pt}{0.600pt}}
\multiput(1310.00,405.50)(0.779,-0.500){157}{\rule{1.083pt}{0.120pt}}
\multiput(1310.00,405.75)(123.751,-81.000){2}{\rule{0.542pt}{0.600pt}}
\sbox{\plotpoint}{\rule[-0.250pt]{0.500pt}{0.500pt}}%
\put(176,566){\raisebox{-.8pt}{\makebox(0,0){$\Diamond$}}}
\put(302,574){\raisebox{-.8pt}{\makebox(0,0){$\Diamond$}}}
\put(428,589){\raisebox{-.8pt}{\makebox(0,0){$\Diamond$}}}
\put(554,614){\raisebox{-.8pt}{\makebox(0,0){$\Diamond$}}}
\put(680,642){\raisebox{-.8pt}{\makebox(0,0){$\Diamond$}}}
\put(806,668){\raisebox{-.8pt}{\makebox(0,0){$\Diamond$}}}
\put(932,629){\raisebox{-.8pt}{\makebox(0,0){$\Diamond$}}}
\put(1058,533){\raisebox{-.8pt}{\makebox(0,0){$\Diamond$}}}
\put(1184,428){\raisebox{-.8pt}{\makebox(0,0){$\Diamond$}}}
\put(1310,335){\raisebox{-.8pt}{\makebox(0,0){$\Diamond$}}}
\put(1436,256){\raisebox{-.8pt}{\makebox(0,0){$\Diamond$}}}
\end{picture}
\end{figure}
\newpage
\begin{figure}
\setlength{\unitlength}{0.240900pt}
\ifx\plotpoint\undefined\newsavebox{\plotpoint}\fi
\sbox{\plotpoint}{\rule[-0.500pt]{1.000pt}{1.000pt}}%
\begin{picture}(1500,1049)(0,0)
\font\gnuplot=cmr10 at 10pt
\gnuplot
\sbox{\plotpoint}{\rule[-0.500pt]{1.000pt}{1.000pt}}%
\put(176.0,68.0){\rule[-0.500pt]{4.818pt}{1.000pt}}
\put(154,68){\makebox(0,0)[r]{$1e-11$}}
\put(1416.0,68.0){\rule[-0.500pt]{4.818pt}{1.000pt}}
\put(176.0,140.0){\rule[-0.500pt]{2.409pt}{1.000pt}}
\put(1426.0,140.0){\rule[-0.500pt]{2.409pt}{1.000pt}}
\put(176.0,182.0){\rule[-0.500pt]{2.409pt}{1.000pt}}
\put(1426.0,182.0){\rule[-0.500pt]{2.409pt}{1.000pt}}
\put(176.0,212.0){\rule[-0.500pt]{2.409pt}{1.000pt}}
\put(1426.0,212.0){\rule[-0.500pt]{2.409pt}{1.000pt}}
\put(176.0,235.0){\rule[-0.500pt]{2.409pt}{1.000pt}}
\put(1426.0,235.0){\rule[-0.500pt]{2.409pt}{1.000pt}}
\put(176.0,254.0){\rule[-0.500pt]{2.409pt}{1.000pt}}
\put(1426.0,254.0){\rule[-0.500pt]{2.409pt}{1.000pt}}
\put(176.0,270.0){\rule[-0.500pt]{2.409pt}{1.000pt}}
\put(1426.0,270.0){\rule[-0.500pt]{2.409pt}{1.000pt}}
\put(176.0,284.0){\rule[-0.500pt]{2.409pt}{1.000pt}}
\put(1426.0,284.0){\rule[-0.500pt]{2.409pt}{1.000pt}}
\put(176.0,297.0){\rule[-0.500pt]{2.409pt}{1.000pt}}
\put(1426.0,297.0){\rule[-0.500pt]{2.409pt}{1.000pt}}
\put(176.0,307.0){\rule[-0.500pt]{4.818pt}{1.000pt}}
\put(154,307){\makebox(0,0)[r]{$1e-10$}}
\put(1416.0,307.0){\rule[-0.500pt]{4.818pt}{1.000pt}}
\put(176.0,380.0){\rule[-0.500pt]{2.409pt}{1.000pt}}
\put(1426.0,380.0){\rule[-0.500pt]{2.409pt}{1.000pt}}
\put(176.0,422.0){\rule[-0.500pt]{2.409pt}{1.000pt}}
\put(1426.0,422.0){\rule[-0.500pt]{2.409pt}{1.000pt}}
\put(176.0,452.0){\rule[-0.500pt]{2.409pt}{1.000pt}}
\put(1426.0,452.0){\rule[-0.500pt]{2.409pt}{1.000pt}}
\put(176.0,475.0){\rule[-0.500pt]{2.409pt}{1.000pt}}
\put(1426.0,475.0){\rule[-0.500pt]{2.409pt}{1.000pt}}
\put(176.0,494.0){\rule[-0.500pt]{2.409pt}{1.000pt}}
\put(1426.0,494.0){\rule[-0.500pt]{2.409pt}{1.000pt}}
\put(176.0,510.0){\rule[-0.500pt]{2.409pt}{1.000pt}}
\put(1426.0,510.0){\rule[-0.500pt]{2.409pt}{1.000pt}}
\put(176.0,524.0){\rule[-0.500pt]{2.409pt}{1.000pt}}
\put(1426.0,524.0){\rule[-0.500pt]{2.409pt}{1.000pt}}
\put(176.0,536.0){\rule[-0.500pt]{2.409pt}{1.000pt}}
\put(1426.0,536.0){\rule[-0.500pt]{2.409pt}{1.000pt}}
\put(176.0,547.0){\rule[-0.500pt]{4.818pt}{1.000pt}}
\put(154,547){\makebox(0,0)[r]{$1e-09$}}
\put(1416.0,547.0){\rule[-0.500pt]{4.818pt}{1.000pt}}
\put(176.0,619.0){\rule[-0.500pt]{2.409pt}{1.000pt}}
\put(1426.0,619.0){\rule[-0.500pt]{2.409pt}{1.000pt}}
\put(176.0,661.0){\rule[-0.500pt]{2.409pt}{1.000pt}}
\put(1426.0,661.0){\rule[-0.500pt]{2.409pt}{1.000pt}}
\put(176.0,691.0){\rule[-0.500pt]{2.409pt}{1.000pt}}
\put(1426.0,691.0){\rule[-0.500pt]{2.409pt}{1.000pt}}
\put(176.0,714.0){\rule[-0.500pt]{2.409pt}{1.000pt}}
\put(1426.0,714.0){\rule[-0.500pt]{2.409pt}{1.000pt}}
\put(176.0,733.0){\rule[-0.500pt]{2.409pt}{1.000pt}}
\put(1426.0,733.0){\rule[-0.500pt]{2.409pt}{1.000pt}}
\put(176.0,749.0){\rule[-0.500pt]{2.409pt}{1.000pt}}
\put(1426.0,749.0){\rule[-0.500pt]{2.409pt}{1.000pt}}
\put(176.0,763.0){\rule[-0.500pt]{2.409pt}{1.000pt}}
\put(1426.0,763.0){\rule[-0.500pt]{2.409pt}{1.000pt}}
\put(176.0,776.0){\rule[-0.500pt]{2.409pt}{1.000pt}}
\put(1426.0,776.0){\rule[-0.500pt]{2.409pt}{1.000pt}}
\put(176.0,786.0){\rule[-0.500pt]{4.818pt}{1.000pt}}
\put(154,786){\makebox(0,0)[r]{$1e-08$}}
\put(1416.0,786.0){\rule[-0.500pt]{4.818pt}{1.000pt}}
\put(176.0,859.0){\rule[-0.500pt]{2.409pt}{1.000pt}}
\put(1426.0,859.0){\rule[-0.500pt]{2.409pt}{1.000pt}}
\put(176.0,901.0){\rule[-0.500pt]{2.409pt}{1.000pt}}
\put(1426.0,901.0){\rule[-0.500pt]{2.409pt}{1.000pt}}
\put(176.0,931.0){\rule[-0.500pt]{2.409pt}{1.000pt}}
\put(1426.0,931.0){\rule[-0.500pt]{2.409pt}{1.000pt}}
\put(176.0,954.0){\rule[-0.500pt]{2.409pt}{1.000pt}}
\put(1426.0,954.0){\rule[-0.500pt]{2.409pt}{1.000pt}}
\put(176.0,973.0){\rule[-0.500pt]{2.409pt}{1.000pt}}
\put(1426.0,973.0){\rule[-0.500pt]{2.409pt}{1.000pt}}
\put(176.0,989.0){\rule[-0.500pt]{2.409pt}{1.000pt}}
\put(1426.0,989.0){\rule[-0.500pt]{2.409pt}{1.000pt}}
\put(176.0,1003.0){\rule[-0.500pt]{2.409pt}{1.000pt}}
\put(1426.0,1003.0){\rule[-0.500pt]{2.409pt}{1.000pt}}
\put(176.0,1015.0){\rule[-0.500pt]{2.409pt}{1.000pt}}
\put(1426.0,1015.0){\rule[-0.500pt]{2.409pt}{1.000pt}}
\put(176.0,1026.0){\rule[-0.500pt]{4.818pt}{1.000pt}}
\put(154,1026){\makebox(0,0)[r]{$1e-07$}}
\put(1416.0,1026.0){\rule[-0.500pt]{4.818pt}{1.000pt}}
\put(176.0,68.0){\rule[-0.500pt]{1.000pt}{4.818pt}}
\put(176,23){\makebox(0,0){$0.5$}}
\put(176.0,1006.0){\rule[-0.500pt]{1.000pt}{4.818pt}}
\put(302.0,68.0){\rule[-0.500pt]{1.000pt}{4.818pt}}
\put(302,23){\makebox(0,0){$0.55$}}
\put(302.0,1006.0){\rule[-0.500pt]{1.000pt}{4.818pt}}
\put(428.0,68.0){\rule[-0.500pt]{1.000pt}{4.818pt}}
\put(428,23){\makebox(0,0){$0.6$}}
\put(428.0,1006.0){\rule[-0.500pt]{1.000pt}{4.818pt}}
\put(554.0,68.0){\rule[-0.500pt]{1.000pt}{4.818pt}}
\put(554,23){\makebox(0,0){$0.65$}}
\put(554.0,1006.0){\rule[-0.500pt]{1.000pt}{4.818pt}}
\put(680.0,68.0){\rule[-0.500pt]{1.000pt}{4.818pt}}
\put(680,23){\makebox(0,0){$0.7$}}
\put(680.0,1006.0){\rule[-0.500pt]{1.000pt}{4.818pt}}
\put(806.0,68.0){\rule[-0.500pt]{1.000pt}{4.818pt}}
\put(806,23){\makebox(0,0){$0.75$}}
\put(806.0,1006.0){\rule[-0.500pt]{1.000pt}{4.818pt}}
\put(932.0,68.0){\rule[-0.500pt]{1.000pt}{4.818pt}}
\put(932,23){\makebox(0,0){$0.8$}}
\put(932.0,1006.0){\rule[-0.500pt]{1.000pt}{4.818pt}}
\put(1058.0,68.0){\rule[-0.500pt]{1.000pt}{4.818pt}}
\put(1058,23){\makebox(0,0){$0.85$}}
\put(1058.0,1006.0){\rule[-0.500pt]{1.000pt}{4.818pt}}
\put(1184.0,68.0){\rule[-0.500pt]{1.000pt}{4.818pt}}
\put(1184,23){\makebox(0,0){$0.9$}}
\put(1184.0,1006.0){\rule[-0.500pt]{1.000pt}{4.818pt}}
\put(1310.0,68.0){\rule[-0.500pt]{1.000pt}{4.818pt}}
\put(1310,23){\makebox(0,0){$0.95$}}
\put(1310.0,1006.0){\rule[-0.500pt]{1.000pt}{4.818pt}}
\put(1436.0,68.0){\rule[-0.500pt]{1.000pt}{4.818pt}}
\put(1436,23){\makebox(0,0){$1$}}
\put(1436.0,1006.0){\rule[-0.500pt]{1.000pt}{4.818pt}}
\put(176.0,68.0){\rule[-0.500pt]{303.534pt}{1.000pt}}
\put(1436.0,68.0){\rule[-0.500pt]{1.000pt}{230.782pt}}
\put(176.0,1026.0){\rule[-0.500pt]{303.534pt}{1.000pt}}
\put(882,-56){\makebox(0,0)[r]{$M(GeV)$}}
\put(882,-296){\makebox(0,0)[r]{{\bf{Figure 1b}}}}
\put(-75,547){\makebox(0,0)[r]{$({dR \over {dM^2d^2p_Tdy}})$}}
\put(882,254){\makebox(0,0)[r]{$T=200MeV$}}
\put(176.0,68.0){\rule[-0.500pt]{1.000pt}{230.782pt}}
\sbox{\plotpoint}{\rule[-0.300pt]{0.600pt}{0.600pt}}%
\put(176,703){\usebox{\plotpoint}}
\multiput(176.00,704.00)(4.352,0.500){25}{\rule{5.190pt}{0.121pt}}
\multiput(176.00,701.75)(115.228,15.000){2}{\rule{2.595pt}{0.600pt}}
\multiput(302.00,719.00)(2.924,0.500){39}{\rule{3.586pt}{0.121pt}}
\multiput(302.00,716.75)(118.556,22.000){2}{\rule{1.793pt}{0.600pt}}
\multiput(428.00,741.00)(2.283,0.500){51}{\rule{2.850pt}{0.120pt}}
\multiput(428.00,738.75)(120.085,28.000){2}{\rule{1.425pt}{0.600pt}}
\multiput(554.00,769.00)(1.719,0.500){69}{\rule{2.193pt}{0.120pt}}
\multiput(554.00,766.75)(121.448,37.000){2}{\rule{1.097pt}{0.600pt}}
\multiput(680.00,806.00)(2.128,0.500){55}{\rule{2.670pt}{0.120pt}}
\multiput(680.00,803.75)(120.458,30.000){2}{\rule{1.335pt}{0.600pt}}
\multiput(806.00,833.50)(2.128,-0.500){55}{\rule{2.670pt}{0.120pt}}
\multiput(806.00,833.75)(120.458,-30.000){2}{\rule{1.335pt}{0.600pt}}
\multiput(932.00,803.50)(0.693,-0.500){177}{\rule{0.981pt}{0.120pt}}
\multiput(932.00,803.75)(123.964,-91.000){2}{\rule{0.490pt}{0.600pt}}
\multiput(1058.00,712.50)(0.663,-0.500){185}{\rule{0.946pt}{0.120pt}}
\multiput(1058.00,712.75)(124.037,-95.000){2}{\rule{0.473pt}{0.600pt}}
\multiput(1184.00,617.50)(0.733,-0.500){167}{\rule{1.029pt}{0.120pt}}
\multiput(1184.00,617.75)(123.864,-86.000){2}{\rule{0.515pt}{0.600pt}}
\multiput(1310.00,531.50)(0.830,-0.500){147}{\rule{1.145pt}{0.120pt}}
\multiput(1310.00,531.75)(123.624,-76.000){2}{\rule{0.572pt}{0.600pt}}
\sbox{\plotpoint}{\rule[-0.250pt]{0.500pt}{0.500pt}}%
\put(176,547){\raisebox{-.8pt}{\makebox(0,0){$\Diamond$}}}
\put(302,566){\raisebox{-.8pt}{\makebox(0,0){$\Diamond$}}}
\put(428,589){\raisebox{-.8pt}{\makebox(0,0){$\Diamond$}}}
\put(554,614){\raisebox{-.8pt}{\makebox(0,0){$\Diamond$}}}
\put(680,654){\raisebox{-.8pt}{\makebox(0,0){$\Diamond$}}}
\put(806,683){\raisebox{-.8pt}{\makebox(0,0){$\Diamond$}}}
\put(932,650){\raisebox{-.8pt}{\makebox(0,0){$\Diamond$}}}
\put(1058,557){\raisebox{-.8pt}{\makebox(0,0){$\Diamond$}}}
\put(1184,464){\raisebox{-.8pt}{\makebox(0,0){$\Diamond$}}}
\put(1310,380){\raisebox{-.8pt}{\makebox(0,0){$\Diamond$}}}
\put(1436,303){\raisebox{-.8pt}{\makebox(0,0){$\Diamond$}}}
\end{picture}
\end{figure}
\newpage
\begin{figure}
\setlength{\unitlength}{0.240900pt}
\ifx\plotpoint\undefined\newsavebox{\plotpoint}\fi
\sbox{\plotpoint}{\rule[-0.500pt]{1.000pt}{1.000pt}}%
\begin{picture}(1500,1049)(0,0)
\font\gnuplot=cmr10 at 10pt
\gnuplot
\sbox{\plotpoint}{\rule[-0.500pt]{1.000pt}{1.000pt}}%
\put(176.0,68.0){\rule[-0.500pt]{4.818pt}{1.000pt}}
\put(154,68){\makebox(0,0)[r]{$1e-06$}}
\put(1416.0,68.0){\rule[-0.500pt]{4.818pt}{1.000pt}}
\put(176.0,140.0){\rule[-0.500pt]{2.409pt}{1.000pt}}
\put(1426.0,140.0){\rule[-0.500pt]{2.409pt}{1.000pt}}
\put(176.0,182.0){\rule[-0.500pt]{2.409pt}{1.000pt}}
\put(1426.0,182.0){\rule[-0.500pt]{2.409pt}{1.000pt}}
\put(176.0,212.0){\rule[-0.500pt]{2.409pt}{1.000pt}}
\put(1426.0,212.0){\rule[-0.500pt]{2.409pt}{1.000pt}}
\put(176.0,235.0){\rule[-0.500pt]{2.409pt}{1.000pt}}
\put(1426.0,235.0){\rule[-0.500pt]{2.409pt}{1.000pt}}
\put(176.0,254.0){\rule[-0.500pt]{2.409pt}{1.000pt}}
\put(1426.0,254.0){\rule[-0.500pt]{2.409pt}{1.000pt}}
\put(176.0,270.0){\rule[-0.500pt]{2.409pt}{1.000pt}}
\put(1426.0,270.0){\rule[-0.500pt]{2.409pt}{1.000pt}}
\put(176.0,284.0){\rule[-0.500pt]{2.409pt}{1.000pt}}
\put(1426.0,284.0){\rule[-0.500pt]{2.409pt}{1.000pt}}
\put(176.0,297.0){\rule[-0.500pt]{2.409pt}{1.000pt}}
\put(1426.0,297.0){\rule[-0.500pt]{2.409pt}{1.000pt}}
\put(176.0,307.0){\rule[-0.500pt]{4.818pt}{1.000pt}}
\put(154,307){\makebox(0,0)[r]{$1e-05$}}
\put(1416.0,307.0){\rule[-0.500pt]{4.818pt}{1.000pt}}
\put(176.0,380.0){\rule[-0.500pt]{2.409pt}{1.000pt}}
\put(1426.0,380.0){\rule[-0.500pt]{2.409pt}{1.000pt}}
\put(176.0,422.0){\rule[-0.500pt]{2.409pt}{1.000pt}}
\put(1426.0,422.0){\rule[-0.500pt]{2.409pt}{1.000pt}}
\put(176.0,452.0){\rule[-0.500pt]{2.409pt}{1.000pt}}
\put(1426.0,452.0){\rule[-0.500pt]{2.409pt}{1.000pt}}
\put(176.0,475.0){\rule[-0.500pt]{2.409pt}{1.000pt}}
\put(1426.0,475.0){\rule[-0.500pt]{2.409pt}{1.000pt}}
\put(176.0,494.0){\rule[-0.500pt]{2.409pt}{1.000pt}}
\put(1426.0,494.0){\rule[-0.500pt]{2.409pt}{1.000pt}}
\put(176.0,510.0){\rule[-0.500pt]{2.409pt}{1.000pt}}
\put(1426.0,510.0){\rule[-0.500pt]{2.409pt}{1.000pt}}
\put(176.0,524.0){\rule[-0.500pt]{2.409pt}{1.000pt}}
\put(1426.0,524.0){\rule[-0.500pt]{2.409pt}{1.000pt}}
\put(176.0,536.0){\rule[-0.500pt]{2.409pt}{1.000pt}}
\put(1426.0,536.0){\rule[-0.500pt]{2.409pt}{1.000pt}}
\put(176.0,547.0){\rule[-0.500pt]{4.818pt}{1.000pt}}
\put(154,547){\makebox(0,0)[r]{$0.0001$}}
\put(1416.0,547.0){\rule[-0.500pt]{4.818pt}{1.000pt}}
\put(176.0,619.0){\rule[-0.500pt]{2.409pt}{1.000pt}}
\put(1426.0,619.0){\rule[-0.500pt]{2.409pt}{1.000pt}}
\put(176.0,661.0){\rule[-0.500pt]{2.409pt}{1.000pt}}
\put(1426.0,661.0){\rule[-0.500pt]{2.409pt}{1.000pt}}
\put(176.0,691.0){\rule[-0.500pt]{2.409pt}{1.000pt}}
\put(1426.0,691.0){\rule[-0.500pt]{2.409pt}{1.000pt}}
\put(176.0,714.0){\rule[-0.500pt]{2.409pt}{1.000pt}}
\put(1426.0,714.0){\rule[-0.500pt]{2.409pt}{1.000pt}}
\put(176.0,733.0){\rule[-0.500pt]{2.409pt}{1.000pt}}
\put(1426.0,733.0){\rule[-0.500pt]{2.409pt}{1.000pt}}
\put(176.0,749.0){\rule[-0.500pt]{2.409pt}{1.000pt}}
\put(1426.0,749.0){\rule[-0.500pt]{2.409pt}{1.000pt}}
\put(176.0,763.0){\rule[-0.500pt]{2.409pt}{1.000pt}}
\put(1426.0,763.0){\rule[-0.500pt]{2.409pt}{1.000pt}}
\put(176.0,776.0){\rule[-0.500pt]{2.409pt}{1.000pt}}
\put(1426.0,776.0){\rule[-0.500pt]{2.409pt}{1.000pt}}
\put(176.0,786.0){\rule[-0.500pt]{4.818pt}{1.000pt}}
\put(154,786){\makebox(0,0)[r]{$0.001$}}
\put(1416.0,786.0){\rule[-0.500pt]{4.818pt}{1.000pt}}
\put(176.0,859.0){\rule[-0.500pt]{2.409pt}{1.000pt}}
\put(1426.0,859.0){\rule[-0.500pt]{2.409pt}{1.000pt}}
\put(176.0,901.0){\rule[-0.500pt]{2.409pt}{1.000pt}}
\put(1426.0,901.0){\rule[-0.500pt]{2.409pt}{1.000pt}}
\put(176.0,931.0){\rule[-0.500pt]{2.409pt}{1.000pt}}
\put(1426.0,931.0){\rule[-0.500pt]{2.409pt}{1.000pt}}
\put(176.0,954.0){\rule[-0.500pt]{2.409pt}{1.000pt}}
\put(1426.0,954.0){\rule[-0.500pt]{2.409pt}{1.000pt}}
\put(176.0,973.0){\rule[-0.500pt]{2.409pt}{1.000pt}}
\put(1426.0,973.0){\rule[-0.500pt]{2.409pt}{1.000pt}}
\put(176.0,989.0){\rule[-0.500pt]{2.409pt}{1.000pt}}
\put(1426.0,989.0){\rule[-0.500pt]{2.409pt}{1.000pt}}
\put(176.0,1003.0){\rule[-0.500pt]{2.409pt}{1.000pt}}
\put(1426.0,1003.0){\rule[-0.500pt]{2.409pt}{1.000pt}}
\put(176.0,1015.0){\rule[-0.500pt]{2.409pt}{1.000pt}}
\put(1426.0,1015.0){\rule[-0.500pt]{2.409pt}{1.000pt}}
\put(176.0,1026.0){\rule[-0.500pt]{4.818pt}{1.000pt}}
\put(154,1026){\makebox(0,0)[r]{$0.01$}}
\put(1416.0,1026.0){\rule[-0.500pt]{4.818pt}{1.000pt}}
\put(176.0,68.0){\rule[-0.500pt]{1.000pt}{4.818pt}}
\put(176,23){\makebox(0,0){$0.4$}}
\put(176.0,1006.0){\rule[-0.500pt]{1.000pt}{4.818pt}}
\put(386.0,68.0){\rule[-0.500pt]{1.000pt}{4.818pt}}
\put(386,23){\makebox(0,0){$0.5$}}
\put(386.0,1006.0){\rule[-0.500pt]{1.000pt}{4.818pt}}
\put(596.0,68.0){\rule[-0.500pt]{1.000pt}{4.818pt}}
\put(596,23){\makebox(0,0){$0.6$}}
\put(596.0,1006.0){\rule[-0.500pt]{1.000pt}{4.818pt}}
\put(806.0,68.0){\rule[-0.500pt]{1.000pt}{4.818pt}}
\put(806,23){\makebox(0,0){$0.7$}}
\put(806.0,1006.0){\rule[-0.500pt]{1.000pt}{4.818pt}}
\put(1016.0,68.0){\rule[-0.500pt]{1.000pt}{4.818pt}}
\put(1016,23){\makebox(0,0){$0.8$}}
\put(1016.0,1006.0){\rule[-0.500pt]{1.000pt}{4.818pt}}
\put(1226.0,68.0){\rule[-0.500pt]{1.000pt}{4.818pt}}
\put(1226,23){\makebox(0,0){$0.9$}}
\put(1226.0,1006.0){\rule[-0.500pt]{1.000pt}{4.818pt}}
\put(1436.0,68.0){\rule[-0.500pt]{1.000pt}{4.818pt}}
\put(1436,23){\makebox(0,0){$1$}}
\put(1436.0,1006.0){\rule[-0.500pt]{1.000pt}{4.818pt}}
\put(176.0,68.0){\rule[-0.500pt]{303.534pt}{1.000pt}}
\put(1436.0,68.0){\rule[-0.500pt]{1.000pt}{230.782pt}}
\put(176.0,1026.0){\rule[-0.500pt]{303.534pt}{1.000pt}}
\put(848,-80){\makebox(0,0)[r]{$M(GeV)$}}
\put(848,-355){\makebox(0,0)[r]{{\bf{Figure 2a}}}}
\put(-25,547){\makebox(0,0)[r]{$({dN \over {dM^2dy}})$}}
\put(848,308){\makebox(0,0)[r]{$T_c=160MeV$}}
\put(176.0,68.0){\rule[-0.500pt]{1.000pt}{230.782pt}}
\sbox{\plotpoint}{\rule[-0.300pt]{0.600pt}{0.600pt}}%
\put(176,884){\usebox{\plotpoint}}
\put(197,882.25){\rule{5.059pt}{0.600pt}}
\multiput(197.00,882.75)(10.500,-1.000){2}{\rule{2.529pt}{0.600pt}}
\put(218,881.25){\rule{5.059pt}{0.600pt}}
\multiput(218.00,881.75)(10.500,-1.000){2}{\rule{2.529pt}{0.600pt}}
\put(239,880.25){\rule{5.059pt}{0.600pt}}
\multiput(239.00,880.75)(10.500,-1.000){2}{\rule{2.529pt}{0.600pt}}
\put(260,878.75){\rule{5.059pt}{0.600pt}}
\multiput(260.00,879.75)(10.500,-2.000){2}{\rule{2.529pt}{0.600pt}}
\put(281,877.25){\rule{5.059pt}{0.600pt}}
\multiput(281.00,877.75)(10.500,-1.000){2}{\rule{2.529pt}{0.600pt}}
\put(302,875.75){\rule{5.059pt}{0.600pt}}
\multiput(302.00,876.75)(10.500,-2.000){2}{\rule{2.529pt}{0.600pt}}
\put(323,873.75){\rule{5.059pt}{0.600pt}}
\multiput(323.00,874.75)(10.500,-2.000){2}{\rule{2.529pt}{0.600pt}}
\put(344,871.75){\rule{5.059pt}{0.600pt}}
\multiput(344.00,872.75)(10.500,-2.000){2}{\rule{2.529pt}{0.600pt}}
\put(365,870.25){\rule{5.059pt}{0.600pt}}
\multiput(365.00,870.75)(10.500,-1.000){2}{\rule{2.529pt}{0.600pt}}
\put(386,868.75){\rule{5.059pt}{0.600pt}}
\multiput(386.00,869.75)(10.500,-2.000){2}{\rule{2.529pt}{0.600pt}}
\put(407,867.25){\rule{5.059pt}{0.600pt}}
\multiput(407.00,867.75)(10.500,-1.000){2}{\rule{2.529pt}{0.600pt}}
\put(428,865.75){\rule{5.059pt}{0.600pt}}
\multiput(428.00,866.75)(10.500,-2.000){2}{\rule{2.529pt}{0.600pt}}
\put(449,864.25){\rule{5.059pt}{0.600pt}}
\multiput(449.00,864.75)(10.500,-1.000){2}{\rule{2.529pt}{0.600pt}}
\put(470,863.25){\rule{5.059pt}{0.600pt}}
\multiput(470.00,863.75)(10.500,-1.000){2}{\rule{2.529pt}{0.600pt}}
\put(176.0,884.0){\rule[-0.300pt]{5.059pt}{0.600pt}}
\put(512,862.25){\rule{5.059pt}{0.600pt}}
\multiput(512.00,862.75)(10.500,-1.000){2}{\rule{2.529pt}{0.600pt}}
\put(491.0,864.0){\rule[-0.300pt]{5.059pt}{0.600pt}}
\put(554,862.25){\rule{5.059pt}{0.600pt}}
\multiput(554.00,861.75)(10.500,1.000){2}{\rule{2.529pt}{0.600pt}}
\put(533.0,863.0){\rule[-0.300pt]{5.059pt}{0.600pt}}
\put(596,863.25){\rule{5.059pt}{0.600pt}}
\multiput(596.00,862.75)(10.500,1.000){2}{\rule{2.529pt}{0.600pt}}
\put(617,864.75){\rule{5.059pt}{0.600pt}}
\multiput(617.00,863.75)(10.500,2.000){2}{\rule{2.529pt}{0.600pt}}
\put(638,866.75){\rule{5.059pt}{0.600pt}}
\multiput(638.00,865.75)(10.500,2.000){2}{\rule{2.529pt}{0.600pt}}
\put(659,868.75){\rule{5.059pt}{0.600pt}}
\multiput(659.00,867.75)(10.500,2.000){2}{\rule{2.529pt}{0.600pt}}
\put(680,871.25){\rule{4.350pt}{0.600pt}}
\multiput(680.00,869.75)(11.971,3.000){2}{\rule{2.175pt}{0.600pt}}
\put(701,874.25){\rule{4.350pt}{0.600pt}}
\multiput(701.00,872.75)(11.971,3.000){2}{\rule{2.175pt}{0.600pt}}
\put(722,877.25){\rule{4.350pt}{0.600pt}}
\multiput(722.00,875.75)(11.971,3.000){2}{\rule{2.175pt}{0.600pt}}
\multiput(743.00,880.99)(3.651,0.503){3}{\rule{3.300pt}{0.121pt}}
\multiput(743.00,878.75)(14.151,4.000){2}{\rule{1.650pt}{0.600pt}}
\multiput(764.00,884.99)(2.479,0.502){5}{\rule{2.670pt}{0.121pt}}
\multiput(764.00,882.75)(15.458,5.000){2}{\rule{1.335pt}{0.600pt}}
\multiput(785.00,889.99)(3.651,0.503){3}{\rule{3.300pt}{0.121pt}}
\multiput(785.00,887.75)(14.151,4.000){2}{\rule{1.650pt}{0.600pt}}
\multiput(806.00,893.99)(2.479,0.502){5}{\rule{2.670pt}{0.121pt}}
\multiput(806.00,891.75)(15.458,5.000){2}{\rule{1.335pt}{0.600pt}}
\multiput(827.00,898.99)(3.651,0.503){3}{\rule{3.300pt}{0.121pt}}
\multiput(827.00,896.75)(14.151,4.000){2}{\rule{1.650pt}{0.600pt}}
\put(848,902.25){\rule{4.350pt}{0.600pt}}
\multiput(848.00,900.75)(11.971,3.000){2}{\rule{2.175pt}{0.600pt}}
\put(869,904.75){\rule{5.059pt}{0.600pt}}
\multiput(869.00,903.75)(10.500,2.000){2}{\rule{2.529pt}{0.600pt}}
\put(575.0,864.0){\rule[-0.300pt]{5.059pt}{0.600pt}}
\put(911,904.25){\rule{4.350pt}{0.600pt}}
\multiput(911.00,905.75)(11.971,-3.000){2}{\rule{2.175pt}{0.600pt}}
\multiput(932.00,902.50)(1.943,-0.501){7}{\rule{2.250pt}{0.121pt}}
\multiput(932.00,902.75)(16.330,-6.000){2}{\rule{1.125pt}{0.600pt}}
\multiput(953.00,896.50)(1.214,-0.501){13}{\rule{1.550pt}{0.121pt}}
\multiput(953.00,896.75)(17.783,-9.000){2}{\rule{0.775pt}{0.600pt}}
\multiput(974.00,887.50)(0.818,-0.500){21}{\rule{1.119pt}{0.121pt}}
\multiput(974.00,887.75)(18.677,-13.000){2}{\rule{0.560pt}{0.600pt}}
\multiput(995.00,874.50)(0.658,-0.500){27}{\rule{0.938pt}{0.121pt}}
\multiput(995.00,874.75)(19.054,-16.000){2}{\rule{0.469pt}{0.600pt}}
\multiput(1016.00,858.50)(0.551,-0.500){33}{\rule{0.813pt}{0.121pt}}
\multiput(1016.00,858.75)(19.312,-19.000){2}{\rule{0.407pt}{0.600pt}}
\multiput(1037.00,839.50)(0.522,-0.500){35}{\rule{0.780pt}{0.121pt}}
\multiput(1037.00,839.75)(19.381,-20.000){2}{\rule{0.390pt}{0.600pt}}
\multiput(1059.00,817.65)(0.500,-0.546){37}{\rule{0.121pt}{0.807pt}}
\multiput(1056.75,819.32)(21.000,-21.325){2}{\rule{0.600pt}{0.404pt}}
\multiput(1080.00,794.77)(0.500,-0.521){37}{\rule{0.121pt}{0.779pt}}
\multiput(1077.75,796.38)(21.000,-20.384){2}{\rule{0.600pt}{0.389pt}}
\multiput(1101.00,772.77)(0.500,-0.521){37}{\rule{0.121pt}{0.779pt}}
\multiput(1098.75,774.38)(21.000,-20.384){2}{\rule{0.600pt}{0.389pt}}
\multiput(1122.00,750.65)(0.500,-0.546){37}{\rule{0.121pt}{0.807pt}}
\multiput(1119.75,752.32)(21.000,-21.325){2}{\rule{0.600pt}{0.404pt}}
\multiput(1143.00,727.77)(0.500,-0.521){37}{\rule{0.121pt}{0.779pt}}
\multiput(1140.75,729.38)(21.000,-20.384){2}{\rule{0.600pt}{0.389pt}}
\multiput(1164.00,705.77)(0.500,-0.521){37}{\rule{0.121pt}{0.779pt}}
\multiput(1161.75,707.38)(21.000,-20.384){2}{\rule{0.600pt}{0.389pt}}
\multiput(1185.00,683.77)(0.500,-0.521){37}{\rule{0.121pt}{0.779pt}}
\multiput(1182.75,685.38)(21.000,-20.384){2}{\rule{0.600pt}{0.389pt}}
\multiput(1205.00,663.50)(0.497,-0.500){37}{\rule{0.750pt}{0.121pt}}
\multiput(1205.00,663.75)(19.443,-21.000){2}{\rule{0.375pt}{0.600pt}}
\multiput(1226.00,642.50)(0.497,-0.500){37}{\rule{0.750pt}{0.121pt}}
\multiput(1226.00,642.75)(19.443,-21.000){2}{\rule{0.375pt}{0.600pt}}
\multiput(1247.00,621.50)(0.522,-0.500){35}{\rule{0.780pt}{0.121pt}}
\multiput(1247.00,621.75)(19.381,-20.000){2}{\rule{0.390pt}{0.600pt}}
\multiput(1268.00,601.50)(0.522,-0.500){35}{\rule{0.780pt}{0.121pt}}
\multiput(1268.00,601.75)(19.381,-20.000){2}{\rule{0.390pt}{0.600pt}}
\multiput(1289.00,581.50)(0.551,-0.500){33}{\rule{0.813pt}{0.121pt}}
\multiput(1289.00,581.75)(19.312,-19.000){2}{\rule{0.407pt}{0.600pt}}
\multiput(1310.00,562.50)(0.551,-0.500){33}{\rule{0.813pt}{0.121pt}}
\multiput(1310.00,562.75)(19.312,-19.000){2}{\rule{0.407pt}{0.600pt}}
\multiput(1331.00,543.50)(0.551,-0.500){33}{\rule{0.813pt}{0.121pt}}
\multiput(1331.00,543.75)(19.312,-19.000){2}{\rule{0.407pt}{0.600pt}}
\multiput(1352.00,524.50)(0.582,-0.500){31}{\rule{0.850pt}{0.121pt}}
\multiput(1352.00,524.75)(19.236,-18.000){2}{\rule{0.425pt}{0.600pt}}
\multiput(1373.00,506.50)(0.618,-0.500){29}{\rule{0.891pt}{0.121pt}}
\multiput(1373.00,506.75)(19.150,-17.000){2}{\rule{0.446pt}{0.600pt}}
\multiput(1394.00,489.50)(0.582,-0.500){31}{\rule{0.850pt}{0.121pt}}
\multiput(1394.00,489.75)(19.236,-18.000){2}{\rule{0.425pt}{0.600pt}}
\multiput(1415.00,471.50)(0.618,-0.500){29}{\rule{0.891pt}{0.121pt}}
\multiput(1415.00,471.75)(19.150,-17.000){2}{\rule{0.446pt}{0.600pt}}
\put(890.0,907.0){\rule[-0.300pt]{5.059pt}{0.600pt}}
\sbox{\plotpoint}{\rule[-0.250pt]{0.500pt}{0.500pt}}%
\put(176,862){\raisebox{-.8pt}{\makebox(0,0){$\Diamond$}}}
\put(197,861){\raisebox{-.8pt}{\makebox(0,0){$\Diamond$}}}
\put(218,860){\raisebox{-.8pt}{\makebox(0,0){$\Diamond$}}}
\put(239,859){\raisebox{-.8pt}{\makebox(0,0){$\Diamond$}}}
\put(260,857){\raisebox{-.8pt}{\makebox(0,0){$\Diamond$}}}
\put(281,855){\raisebox{-.8pt}{\makebox(0,0){$\Diamond$}}}
\put(302,853){\raisebox{-.8pt}{\makebox(0,0){$\Diamond$}}}
\put(323,851){\raisebox{-.8pt}{\makebox(0,0){$\Diamond$}}}
\put(344,849){\raisebox{-.8pt}{\makebox(0,0){$\Diamond$}}}
\put(365,846){\raisebox{-.8pt}{\makebox(0,0){$\Diamond$}}}
\put(386,844){\raisebox{-.8pt}{\makebox(0,0){$\Diamond$}}}
\put(407,842){\raisebox{-.8pt}{\makebox(0,0){$\Diamond$}}}
\put(428,840){\raisebox{-.8pt}{\makebox(0,0){$\Diamond$}}}
\put(449,839){\raisebox{-.8pt}{\makebox(0,0){$\Diamond$}}}
\put(470,837){\raisebox{-.8pt}{\makebox(0,0){$\Diamond$}}}
\put(491,835){\raisebox{-.8pt}{\makebox(0,0){$\Diamond$}}}
\put(512,835){\raisebox{-.8pt}{\makebox(0,0){$\Diamond$}}}
\put(533,834){\raisebox{-.8pt}{\makebox(0,0){$\Diamond$}}}
\put(554,833){\raisebox{-.8pt}{\makebox(0,0){$\Diamond$}}}
\put(575,833){\raisebox{-.8pt}{\makebox(0,0){$\Diamond$}}}
\put(596,833){\raisebox{-.8pt}{\makebox(0,0){$\Diamond$}}}
\put(617,834){\raisebox{-.8pt}{\makebox(0,0){$\Diamond$}}}
\put(638,835){\raisebox{-.8pt}{\makebox(0,0){$\Diamond$}}}
\put(659,837){\raisebox{-.8pt}{\makebox(0,0){$\Diamond$}}}
\put(680,839){\raisebox{-.8pt}{\makebox(0,0){$\Diamond$}}}
\put(701,841){\raisebox{-.8pt}{\makebox(0,0){$\Diamond$}}}
\put(722,844){\raisebox{-.8pt}{\makebox(0,0){$\Diamond$}}}
\put(743,847){\raisebox{-.8pt}{\makebox(0,0){$\Diamond$}}}
\put(764,851){\raisebox{-.8pt}{\makebox(0,0){$\Diamond$}}}
\put(785,854){\raisebox{-.8pt}{\makebox(0,0){$\Diamond$}}}
\put(806,859){\raisebox{-.8pt}{\makebox(0,0){$\Diamond$}}}
\put(827,863){\raisebox{-.8pt}{\makebox(0,0){$\Diamond$}}}
\put(848,867){\raisebox{-.8pt}{\makebox(0,0){$\Diamond$}}}
\put(869,870){\raisebox{-.8pt}{\makebox(0,0){$\Diamond$}}}
\put(890,871){\raisebox{-.8pt}{\makebox(0,0){$\Diamond$}}}
\put(911,871){\raisebox{-.8pt}{\makebox(0,0){$\Diamond$}}}
\put(932,868){\raisebox{-.8pt}{\makebox(0,0){$\Diamond$}}}
\put(953,862){\raisebox{-.8pt}{\makebox(0,0){$\Diamond$}}}
\put(974,852){\raisebox{-.8pt}{\makebox(0,0){$\Diamond$}}}
\put(995,839){\raisebox{-.8pt}{\makebox(0,0){$\Diamond$}}}
\put(1016,822){\raisebox{-.8pt}{\makebox(0,0){$\Diamond$}}}
\put(1037,803){\raisebox{-.8pt}{\makebox(0,0){$\Diamond$}}}
\put(1058,782){\raisebox{-.8pt}{\makebox(0,0){$\Diamond$}}}
\put(1079,760){\raisebox{-.8pt}{\makebox(0,0){$\Diamond$}}}
\put(1100,738){\raisebox{-.8pt}{\makebox(0,0){$\Diamond$}}}
\put(1121,715){\raisebox{-.8pt}{\makebox(0,0){$\Diamond$}}}
\put(1142,692){\raisebox{-.8pt}{\makebox(0,0){$\Diamond$}}}
\put(1163,669){\raisebox{-.8pt}{\makebox(0,0){$\Diamond$}}}
\put(1184,647){\raisebox{-.8pt}{\makebox(0,0){$\Diamond$}}}
\put(1205,625){\raisebox{-.8pt}{\makebox(0,0){$\Diamond$}}}
\put(1226,603){\raisebox{-.8pt}{\makebox(0,0){$\Diamond$}}}
\put(1247,583){\raisebox{-.8pt}{\makebox(0,0){$\Diamond$}}}
\put(1268,562){\raisebox{-.8pt}{\makebox(0,0){$\Diamond$}}}
\put(1289,542){\raisebox{-.8pt}{\makebox(0,0){$\Diamond$}}}
\put(1310,522){\raisebox{-.8pt}{\makebox(0,0){$\Diamond$}}}
\put(1331,503){\raisebox{-.8pt}{\makebox(0,0){$\Diamond$}}}
\put(1352,485){\raisebox{-.8pt}{\makebox(0,0){$\Diamond$}}}
\put(1373,466){\raisebox{-.8pt}{\makebox(0,0){$\Diamond$}}}
\put(1394,448){\raisebox{-.8pt}{\makebox(0,0){$\Diamond$}}}
\put(1415,431){\raisebox{-.8pt}{\makebox(0,0){$\Diamond$}}}
\put(1436,413){\raisebox{-.8pt}{\makebox(0,0){$\Diamond$}}}
\end{picture}
\end{figure}
\newpage
\begin{figure}
\setlength{\unitlength}{0.240900pt}
\ifx\plotpoint\undefined\newsavebox{\plotpoint}\fi
\sbox{\plotpoint}{\rule[-0.500pt]{1.000pt}{1.000pt}}%
\begin{picture}(1500,1049)(0,0)
\font\gnuplot=cmr10 at 10pt
\gnuplot
\sbox{\plotpoint}{\rule[-0.500pt]{1.000pt}{1.000pt}}%
\put(176.0,68.0){\rule[-0.500pt]{4.818pt}{1.000pt}}
\put(154,68){\makebox(0,0)[r]{$1e-06$}}
\put(1416.0,68.0){\rule[-0.500pt]{4.818pt}{1.000pt}}
\put(176.0,140.0){\rule[-0.500pt]{2.409pt}{1.000pt}}
\put(1426.0,140.0){\rule[-0.500pt]{2.409pt}{1.000pt}}
\put(176.0,182.0){\rule[-0.500pt]{2.409pt}{1.000pt}}
\put(1426.0,182.0){\rule[-0.500pt]{2.409pt}{1.000pt}}
\put(176.0,212.0){\rule[-0.500pt]{2.409pt}{1.000pt}}
\put(1426.0,212.0){\rule[-0.500pt]{2.409pt}{1.000pt}}
\put(176.0,235.0){\rule[-0.500pt]{2.409pt}{1.000pt}}
\put(1426.0,235.0){\rule[-0.500pt]{2.409pt}{1.000pt}}
\put(176.0,254.0){\rule[-0.500pt]{2.409pt}{1.000pt}}
\put(1426.0,254.0){\rule[-0.500pt]{2.409pt}{1.000pt}}
\put(176.0,270.0){\rule[-0.500pt]{2.409pt}{1.000pt}}
\put(1426.0,270.0){\rule[-0.500pt]{2.409pt}{1.000pt}}
\put(176.0,284.0){\rule[-0.500pt]{2.409pt}{1.000pt}}
\put(1426.0,284.0){\rule[-0.500pt]{2.409pt}{1.000pt}}
\put(176.0,297.0){\rule[-0.500pt]{2.409pt}{1.000pt}}
\put(1426.0,297.0){\rule[-0.500pt]{2.409pt}{1.000pt}}
\put(176.0,307.0){\rule[-0.500pt]{4.818pt}{1.000pt}}
\put(154,307){\makebox(0,0)[r]{$1e-05$}}
\put(1416.0,307.0){\rule[-0.500pt]{4.818pt}{1.000pt}}
\put(176.0,380.0){\rule[-0.500pt]{2.409pt}{1.000pt}}
\put(1426.0,380.0){\rule[-0.500pt]{2.409pt}{1.000pt}}
\put(176.0,422.0){\rule[-0.500pt]{2.409pt}{1.000pt}}
\put(1426.0,422.0){\rule[-0.500pt]{2.409pt}{1.000pt}}
\put(176.0,452.0){\rule[-0.500pt]{2.409pt}{1.000pt}}
\put(1426.0,452.0){\rule[-0.500pt]{2.409pt}{1.000pt}}
\put(176.0,475.0){\rule[-0.500pt]{2.409pt}{1.000pt}}
\put(1426.0,475.0){\rule[-0.500pt]{2.409pt}{1.000pt}}
\put(176.0,494.0){\rule[-0.500pt]{2.409pt}{1.000pt}}
\put(1426.0,494.0){\rule[-0.500pt]{2.409pt}{1.000pt}}
\put(176.0,510.0){\rule[-0.500pt]{2.409pt}{1.000pt}}
\put(1426.0,510.0){\rule[-0.500pt]{2.409pt}{1.000pt}}
\put(176.0,524.0){\rule[-0.500pt]{2.409pt}{1.000pt}}
\put(1426.0,524.0){\rule[-0.500pt]{2.409pt}{1.000pt}}
\put(176.0,536.0){\rule[-0.500pt]{2.409pt}{1.000pt}}
\put(1426.0,536.0){\rule[-0.500pt]{2.409pt}{1.000pt}}
\put(176.0,547.0){\rule[-0.500pt]{4.818pt}{1.000pt}}
\put(154,547){\makebox(0,0)[r]{$0.0001$}}
\put(1416.0,547.0){\rule[-0.500pt]{4.818pt}{1.000pt}}
\put(176.0,619.0){\rule[-0.500pt]{2.409pt}{1.000pt}}
\put(1426.0,619.0){\rule[-0.500pt]{2.409pt}{1.000pt}}
\put(176.0,661.0){\rule[-0.500pt]{2.409pt}{1.000pt}}
\put(1426.0,661.0){\rule[-0.500pt]{2.409pt}{1.000pt}}
\put(176.0,691.0){\rule[-0.500pt]{2.409pt}{1.000pt}}
\put(1426.0,691.0){\rule[-0.500pt]{2.409pt}{1.000pt}}
\put(176.0,714.0){\rule[-0.500pt]{2.409pt}{1.000pt}}
\put(1426.0,714.0){\rule[-0.500pt]{2.409pt}{1.000pt}}
\put(176.0,733.0){\rule[-0.500pt]{2.409pt}{1.000pt}}
\put(1426.0,733.0){\rule[-0.500pt]{2.409pt}{1.000pt}}
\put(176.0,749.0){\rule[-0.500pt]{2.409pt}{1.000pt}}
\put(1426.0,749.0){\rule[-0.500pt]{2.409pt}{1.000pt}}
\put(176.0,763.0){\rule[-0.500pt]{2.409pt}{1.000pt}}
\put(1426.0,763.0){\rule[-0.500pt]{2.409pt}{1.000pt}}
\put(176.0,776.0){\rule[-0.500pt]{2.409pt}{1.000pt}}
\put(1426.0,776.0){\rule[-0.500pt]{2.409pt}{1.000pt}}
\put(176.0,786.0){\rule[-0.500pt]{4.818pt}{1.000pt}}
\put(154,786){\makebox(0,0)[r]{$0.001$}}
\put(1416.0,786.0){\rule[-0.500pt]{4.818pt}{1.000pt}}
\put(176.0,859.0){\rule[-0.500pt]{2.409pt}{1.000pt}}
\put(1426.0,859.0){\rule[-0.500pt]{2.409pt}{1.000pt}}
\put(176.0,901.0){\rule[-0.500pt]{2.409pt}{1.000pt}}
\put(1426.0,901.0){\rule[-0.500pt]{2.409pt}{1.000pt}}
\put(176.0,931.0){\rule[-0.500pt]{2.409pt}{1.000pt}}
\put(1426.0,931.0){\rule[-0.500pt]{2.409pt}{1.000pt}}
\put(176.0,954.0){\rule[-0.500pt]{2.409pt}{1.000pt}}
\put(1426.0,954.0){\rule[-0.500pt]{2.409pt}{1.000pt}}
\put(176.0,973.0){\rule[-0.500pt]{2.409pt}{1.000pt}}
\put(1426.0,973.0){\rule[-0.500pt]{2.409pt}{1.000pt}}
\put(176.0,989.0){\rule[-0.500pt]{2.409pt}{1.000pt}}
\put(1426.0,989.0){\rule[-0.500pt]{2.409pt}{1.000pt}}
\put(176.0,1003.0){\rule[-0.500pt]{2.409pt}{1.000pt}}
\put(1426.0,1003.0){\rule[-0.500pt]{2.409pt}{1.000pt}}
\put(176.0,1015.0){\rule[-0.500pt]{2.409pt}{1.000pt}}
\put(1426.0,1015.0){\rule[-0.500pt]{2.409pt}{1.000pt}}
\put(176.0,1026.0){\rule[-0.500pt]{4.818pt}{1.000pt}}
\put(154,1026){\makebox(0,0)[r]{$0.01$}}
\put(1416.0,1026.0){\rule[-0.500pt]{4.818pt}{1.000pt}}
\put(176.0,68.0){\rule[-0.500pt]{1.000pt}{4.818pt}}
\put(176,23){\makebox(0,0){$0.4$}}
\put(176.0,1006.0){\rule[-0.500pt]{1.000pt}{4.818pt}}
\put(386.0,68.0){\rule[-0.500pt]{1.000pt}{4.818pt}}
\put(386,23){\makebox(0,0){$0.5$}}
\put(386.0,1006.0){\rule[-0.500pt]{1.000pt}{4.818pt}}
\put(596.0,68.0){\rule[-0.500pt]{1.000pt}{4.818pt}}
\put(596,23){\makebox(0,0){$0.6$}}
\put(596.0,1006.0){\rule[-0.500pt]{1.000pt}{4.818pt}}
\put(806.0,68.0){\rule[-0.500pt]{1.000pt}{4.818pt}}
\put(806,23){\makebox(0,0){$0.7$}}
\put(806.0,1006.0){\rule[-0.500pt]{1.000pt}{4.818pt}}
\put(1016.0,68.0){\rule[-0.500pt]{1.000pt}{4.818pt}}
\put(1016,23){\makebox(0,0){$0.8$}}
\put(1016.0,1006.0){\rule[-0.500pt]{1.000pt}{4.818pt}}
\put(1226.0,68.0){\rule[-0.500pt]{1.000pt}{4.818pt}}
\put(1226,23){\makebox(0,0){$0.9$}}
\put(1226.0,1006.0){\rule[-0.500pt]{1.000pt}{4.818pt}}
\put(1436.0,68.0){\rule[-0.500pt]{1.000pt}{4.818pt}}
\put(1436,23){\makebox(0,0){$1$}}
\put(1436.0,1006.0){\rule[-0.500pt]{1.000pt}{4.818pt}}
\put(176.0,68.0){\rule[-0.500pt]{303.534pt}{1.000pt}}
\put(1436.0,68.0){\rule[-0.500pt]{1.000pt}{230.782pt}}
\put(176.0,1026.0){\rule[-0.500pt]{303.534pt}{1.000pt}}
\put(911,-171){\makebox(0,0)[r]{$M(GeV)$}}
\put(911,-410){\makebox(0,0)[r]{{\bf{Figure 2b}}}}
\put(8,547){\makebox(0,0)[r]{$({dN \over {dM^2dy}})$}}
\put(911,308){\makebox(0,0)[r]{$T_c=200MeV$}}
\put(176.0,68.0){\rule[-0.500pt]{1.000pt}{230.782pt}}
\sbox{\plotpoint}{\rule[-0.300pt]{0.600pt}{0.600pt}}%
\put(176,894){\usebox{\plotpoint}}
\put(176,893.25){\rule{5.059pt}{0.600pt}}
\multiput(176.00,892.75)(10.500,1.000){2}{\rule{2.529pt}{0.600pt}}
\put(218,893.25){\rule{5.059pt}{0.600pt}}
\multiput(218.00,893.75)(10.500,-1.000){2}{\rule{2.529pt}{0.600pt}}
\put(197.0,895.0){\rule[-0.300pt]{5.059pt}{0.600pt}}
\put(260,892.25){\rule{5.059pt}{0.600pt}}
\multiput(260.00,892.75)(10.500,-1.000){2}{\rule{2.529pt}{0.600pt}}
\put(281,891.25){\rule{5.059pt}{0.600pt}}
\multiput(281.00,891.75)(10.500,-1.000){2}{\rule{2.529pt}{0.600pt}}
\put(302,890.25){\rule{5.059pt}{0.600pt}}
\multiput(302.00,890.75)(10.500,-1.000){2}{\rule{2.529pt}{0.600pt}}
\put(323,889.25){\rule{5.059pt}{0.600pt}}
\multiput(323.00,889.75)(10.500,-1.000){2}{\rule{2.529pt}{0.600pt}}
\put(344,888.25){\rule{5.059pt}{0.600pt}}
\multiput(344.00,888.75)(10.500,-1.000){2}{\rule{2.529pt}{0.600pt}}
\put(365,887.25){\rule{5.059pt}{0.600pt}}
\multiput(365.00,887.75)(10.500,-1.000){2}{\rule{2.529pt}{0.600pt}}
\put(386,886.25){\rule{5.059pt}{0.600pt}}
\multiput(386.00,886.75)(10.500,-1.000){2}{\rule{2.529pt}{0.600pt}}
\put(407,885.25){\rule{5.059pt}{0.600pt}}
\multiput(407.00,885.75)(10.500,-1.000){2}{\rule{2.529pt}{0.600pt}}
\put(239.0,894.0){\rule[-0.300pt]{5.059pt}{0.600pt}}
\put(533,885.25){\rule{5.059pt}{0.600pt}}
\multiput(533.00,884.75)(10.500,1.000){2}{\rule{2.529pt}{0.600pt}}
\put(554,886.25){\rule{5.059pt}{0.600pt}}
\multiput(554.00,885.75)(10.500,1.000){2}{\rule{2.529pt}{0.600pt}}
\put(575,887.75){\rule{5.059pt}{0.600pt}}
\multiput(575.00,886.75)(10.500,2.000){2}{\rule{2.529pt}{0.600pt}}
\put(596,889.25){\rule{5.059pt}{0.600pt}}
\multiput(596.00,888.75)(10.500,1.000){2}{\rule{2.529pt}{0.600pt}}
\put(617,891.25){\rule{4.350pt}{0.600pt}}
\multiput(617.00,889.75)(11.971,3.000){2}{\rule{2.175pt}{0.600pt}}
\put(638,894.25){\rule{4.350pt}{0.600pt}}
\multiput(638.00,892.75)(11.971,3.000){2}{\rule{2.175pt}{0.600pt}}
\put(659,897.25){\rule{4.350pt}{0.600pt}}
\multiput(659.00,895.75)(11.971,3.000){2}{\rule{2.175pt}{0.600pt}}
\put(680,900.25){\rule{4.350pt}{0.600pt}}
\multiput(680.00,898.75)(11.971,3.000){2}{\rule{2.175pt}{0.600pt}}
\multiput(701.00,903.99)(2.479,0.502){5}{\rule{2.670pt}{0.121pt}}
\multiput(701.00,901.75)(15.458,5.000){2}{\rule{1.335pt}{0.600pt}}
\multiput(722.00,908.99)(3.651,0.503){3}{\rule{3.300pt}{0.121pt}}
\multiput(722.00,906.75)(14.151,4.000){2}{\rule{1.650pt}{0.600pt}}
\multiput(743.00,912.99)(2.479,0.502){5}{\rule{2.670pt}{0.121pt}}
\multiput(743.00,910.75)(15.458,5.000){2}{\rule{1.335pt}{0.600pt}}
\multiput(764.00,917.99)(2.479,0.502){5}{\rule{2.670pt}{0.121pt}}
\multiput(764.00,915.75)(15.458,5.000){2}{\rule{1.335pt}{0.600pt}}
\multiput(785.00,922.99)(1.943,0.501){7}{\rule{2.250pt}{0.121pt}}
\multiput(785.00,920.75)(16.330,6.000){2}{\rule{1.125pt}{0.600pt}}
\multiput(806.00,928.99)(2.479,0.502){5}{\rule{2.670pt}{0.121pt}}
\multiput(806.00,926.75)(15.458,5.000){2}{\rule{1.335pt}{0.600pt}}
\multiput(827.00,933.99)(2.479,0.502){5}{\rule{2.670pt}{0.121pt}}
\multiput(827.00,931.75)(15.458,5.000){2}{\rule{1.335pt}{0.600pt}}
\multiput(848.00,938.99)(2.479,0.502){5}{\rule{2.670pt}{0.121pt}}
\multiput(848.00,936.75)(15.458,5.000){2}{\rule{1.335pt}{0.600pt}}
\put(869,943.25){\rule{4.350pt}{0.600pt}}
\multiput(869.00,941.75)(11.971,3.000){2}{\rule{2.175pt}{0.600pt}}
\put(890,945.25){\rule{5.059pt}{0.600pt}}
\multiput(890.00,944.75)(10.500,1.000){2}{\rule{2.529pt}{0.600pt}}
\put(911,944.75){\rule{5.059pt}{0.600pt}}
\multiput(911.00,945.75)(10.500,-2.000){2}{\rule{2.529pt}{0.600pt}}
\multiput(932.00,943.50)(2.479,-0.502){5}{\rule{2.670pt}{0.121pt}}
\multiput(932.00,943.75)(15.458,-5.000){2}{\rule{1.335pt}{0.600pt}}
\multiput(953.00,938.50)(1.384,-0.501){11}{\rule{1.725pt}{0.121pt}}
\multiput(953.00,938.75)(17.420,-8.000){2}{\rule{0.863pt}{0.600pt}}
\multiput(974.00,930.50)(0.890,-0.500){19}{\rule{1.200pt}{0.121pt}}
\multiput(974.00,930.75)(18.509,-12.000){2}{\rule{0.600pt}{0.600pt}}
\multiput(995.00,918.50)(0.658,-0.500){27}{\rule{0.938pt}{0.121pt}}
\multiput(995.00,918.75)(19.054,-16.000){2}{\rule{0.469pt}{0.600pt}}
\multiput(1016.00,902.50)(0.618,-0.500){29}{\rule{0.891pt}{0.121pt}}
\multiput(1016.00,902.75)(19.150,-17.000){2}{\rule{0.446pt}{0.600pt}}
\multiput(1037.00,885.50)(0.522,-0.500){35}{\rule{0.780pt}{0.121pt}}
\multiput(1037.00,885.75)(19.381,-20.000){2}{\rule{0.390pt}{0.600pt}}
\multiput(1058.00,865.50)(0.522,-0.500){35}{\rule{0.780pt}{0.121pt}}
\multiput(1058.00,865.75)(19.381,-20.000){2}{\rule{0.390pt}{0.600pt}}
\multiput(1080.00,843.77)(0.500,-0.521){37}{\rule{0.121pt}{0.779pt}}
\multiput(1077.75,845.38)(21.000,-20.384){2}{\rule{0.600pt}{0.389pt}}
\multiput(1100.00,823.50)(0.497,-0.500){37}{\rule{0.750pt}{0.121pt}}
\multiput(1100.00,823.75)(19.443,-21.000){2}{\rule{0.375pt}{0.600pt}}
\multiput(1122.00,800.77)(0.500,-0.521){37}{\rule{0.121pt}{0.779pt}}
\multiput(1119.75,802.38)(21.000,-20.384){2}{\rule{0.600pt}{0.389pt}}
\multiput(1142.00,780.50)(0.497,-0.500){37}{\rule{0.750pt}{0.121pt}}
\multiput(1142.00,780.75)(19.443,-21.000){2}{\rule{0.375pt}{0.600pt}}
\multiput(1163.00,759.50)(0.497,-0.500){37}{\rule{0.750pt}{0.121pt}}
\multiput(1163.00,759.75)(19.443,-21.000){2}{\rule{0.375pt}{0.600pt}}
\multiput(1184.00,738.50)(0.522,-0.500){35}{\rule{0.780pt}{0.121pt}}
\multiput(1184.00,738.75)(19.381,-20.000){2}{\rule{0.390pt}{0.600pt}}
\multiput(1205.00,718.50)(0.497,-0.500){37}{\rule{0.750pt}{0.121pt}}
\multiput(1205.00,718.75)(19.443,-21.000){2}{\rule{0.375pt}{0.600pt}}
\multiput(1226.00,697.50)(0.551,-0.500){33}{\rule{0.813pt}{0.121pt}}
\multiput(1226.00,697.75)(19.312,-19.000){2}{\rule{0.407pt}{0.600pt}}
\multiput(1247.00,678.50)(0.551,-0.500){33}{\rule{0.813pt}{0.121pt}}
\multiput(1247.00,678.75)(19.312,-19.000){2}{\rule{0.407pt}{0.600pt}}
\multiput(1268.00,659.50)(0.551,-0.500){33}{\rule{0.813pt}{0.121pt}}
\multiput(1268.00,659.75)(19.312,-19.000){2}{\rule{0.407pt}{0.600pt}}
\multiput(1289.00,640.50)(0.582,-0.500){31}{\rule{0.850pt}{0.121pt}}
\multiput(1289.00,640.75)(19.236,-18.000){2}{\rule{0.425pt}{0.600pt}}
\multiput(1310.00,622.50)(0.582,-0.500){31}{\rule{0.850pt}{0.121pt}}
\multiput(1310.00,622.75)(19.236,-18.000){2}{\rule{0.425pt}{0.600pt}}
\multiput(1331.00,604.50)(0.618,-0.500){29}{\rule{0.891pt}{0.121pt}}
\multiput(1331.00,604.75)(19.150,-17.000){2}{\rule{0.446pt}{0.600pt}}
\multiput(1352.00,587.50)(0.582,-0.500){31}{\rule{0.850pt}{0.121pt}}
\multiput(1352.00,587.75)(19.236,-18.000){2}{\rule{0.425pt}{0.600pt}}
\multiput(1373.00,569.50)(0.658,-0.500){27}{\rule{0.938pt}{0.121pt}}
\multiput(1373.00,569.75)(19.054,-16.000){2}{\rule{0.469pt}{0.600pt}}
\multiput(1394.00,553.50)(0.658,-0.500){27}{\rule{0.938pt}{0.121pt}}
\multiput(1394.00,553.75)(19.054,-16.000){2}{\rule{0.469pt}{0.600pt}}
\multiput(1415.00,537.50)(0.658,-0.500){27}{\rule{0.938pt}{0.121pt}}
\multiput(1415.00,537.75)(19.054,-16.000){2}{\rule{0.469pt}{0.600pt}}
\put(428.0,886.0){\rule[-0.300pt]{25.294pt}{0.600pt}}
\sbox{\plotpoint}{\rule[-0.250pt]{0.500pt}{0.500pt}}%
\put(176,867){\raisebox{-.8pt}{\makebox(0,0){$\Diamond$}}}
\put(197,867){\raisebox{-.8pt}{\makebox(0,0){$\Diamond$}}}
\put(218,866){\raisebox{-.8pt}{\makebox(0,0){$\Diamond$}}}
\put(239,865){\raisebox{-.8pt}{\makebox(0,0){$\Diamond$}}}
\put(260,864){\raisebox{-.8pt}{\makebox(0,0){$\Diamond$}}}
\put(281,862){\raisebox{-.8pt}{\makebox(0,0){$\Diamond$}}}
\put(302,860){\raisebox{-.8pt}{\makebox(0,0){$\Diamond$}}}
\put(323,859){\raisebox{-.8pt}{\makebox(0,0){$\Diamond$}}}
\put(344,857){\raisebox{-.8pt}{\makebox(0,0){$\Diamond$}}}
\put(365,855){\raisebox{-.8pt}{\makebox(0,0){$\Diamond$}}}
\put(386,853){\raisebox{-.8pt}{\makebox(0,0){$\Diamond$}}}
\put(407,852){\raisebox{-.8pt}{\makebox(0,0){$\Diamond$}}}
\put(428,850){\raisebox{-.8pt}{\makebox(0,0){$\Diamond$}}}
\put(449,849){\raisebox{-.8pt}{\makebox(0,0){$\Diamond$}}}
\put(470,848){\raisebox{-.8pt}{\makebox(0,0){$\Diamond$}}}
\put(491,847){\raisebox{-.8pt}{\makebox(0,0){$\Diamond$}}}
\put(512,847){\raisebox{-.8pt}{\makebox(0,0){$\Diamond$}}}
\put(533,846){\raisebox{-.8pt}{\makebox(0,0){$\Diamond$}}}
\put(554,846){\raisebox{-.8pt}{\makebox(0,0){$\Diamond$}}}
\put(575,847){\raisebox{-.8pt}{\makebox(0,0){$\Diamond$}}}
\put(596,847){\raisebox{-.8pt}{\makebox(0,0){$\Diamond$}}}
\put(617,848){\raisebox{-.8pt}{\makebox(0,0){$\Diamond$}}}
\put(638,850){\raisebox{-.8pt}{\makebox(0,0){$\Diamond$}}}
\put(659,852){\raisebox{-.8pt}{\makebox(0,0){$\Diamond$}}}
\put(680,854){\raisebox{-.8pt}{\makebox(0,0){$\Diamond$}}}
\put(701,858){\raisebox{-.8pt}{\makebox(0,0){$\Diamond$}}}
\put(722,861){\raisebox{-.8pt}{\makebox(0,0){$\Diamond$}}}
\put(743,865){\raisebox{-.8pt}{\makebox(0,0){$\Diamond$}}}
\put(764,869){\raisebox{-.8pt}{\makebox(0,0){$\Diamond$}}}
\put(785,874){\raisebox{-.8pt}{\makebox(0,0){$\Diamond$}}}
\put(806,878){\raisebox{-.8pt}{\makebox(0,0){$\Diamond$}}}
\put(827,884){\raisebox{-.8pt}{\makebox(0,0){$\Diamond$}}}
\put(848,888){\raisebox{-.8pt}{\makebox(0,0){$\Diamond$}}}
\put(869,891){\raisebox{-.8pt}{\makebox(0,0){$\Diamond$}}}
\put(890,894){\raisebox{-.8pt}{\makebox(0,0){$\Diamond$}}}
\put(911,894){\raisebox{-.8pt}{\makebox(0,0){$\Diamond$}}}
\put(932,892){\raisebox{-.8pt}{\makebox(0,0){$\Diamond$}}}
\put(953,886){\raisebox{-.8pt}{\makebox(0,0){$\Diamond$}}}
\put(974,877){\raisebox{-.8pt}{\makebox(0,0){$\Diamond$}}}
\put(995,864){\raisebox{-.8pt}{\makebox(0,0){$\Diamond$}}}
\put(1016,849){\raisebox{-.8pt}{\makebox(0,0){$\Diamond$}}}
\put(1037,830){\raisebox{-.8pt}{\makebox(0,0){$\Diamond$}}}
\put(1058,811){\raisebox{-.8pt}{\makebox(0,0){$\Diamond$}}}
\put(1079,789){\raisebox{-.8pt}{\makebox(0,0){$\Diamond$}}}
\put(1100,767){\raisebox{-.8pt}{\makebox(0,0){$\Diamond$}}}
\put(1121,745){\raisebox{-.8pt}{\makebox(0,0){$\Diamond$}}}
\put(1142,723){\raisebox{-.8pt}{\makebox(0,0){$\Diamond$}}}
\put(1163,701){\raisebox{-.8pt}{\makebox(0,0){$\Diamond$}}}
\put(1184,679){\raisebox{-.8pt}{\makebox(0,0){$\Diamond$}}}
\put(1205,658){\raisebox{-.8pt}{\makebox(0,0){$\Diamond$}}}
\put(1226,638){\raisebox{-.8pt}{\makebox(0,0){$\Diamond$}}}
\put(1247,618){\raisebox{-.8pt}{\makebox(0,0){$\Diamond$}}}
\put(1268,598){\raisebox{-.8pt}{\makebox(0,0){$\Diamond$}}}
\put(1289,578){\raisebox{-.8pt}{\makebox(0,0){$\Diamond$}}}
\put(1310,560){\raisebox{-.8pt}{\makebox(0,0){$\Diamond$}}}
\put(1331,541){\raisebox{-.8pt}{\makebox(0,0){$\Diamond$}}}
\put(1352,524){\raisebox{-.8pt}{\makebox(0,0){$\Diamond$}}}
\put(1373,506){\raisebox{-.8pt}{\makebox(0,0){$\Diamond$}}}
\put(1394,489){\raisebox{-.8pt}{\makebox(0,0){$\Diamond$}}}
\put(1415,472){\raisebox{-.8pt}{\makebox(0,0){$\Diamond$}}}
\put(1436,456){\raisebox{-.8pt}{\makebox(0,0){$\Diamond$}}}
\end{picture}
\end{figure}

\newpage
\begin{thebibliography}{200}

\bibitem  {a} E.L.Feinberg, Nuovo Cimento \underline {A34}, 39 (1976).
\bibitem  {b} S.Raha and B.Sinha, Int.J.Mod.Phys. \underline {A6}, 517 
(1991).
\bibitem  {c} J.Alam, S.Raha and B.Sinha, Phys.Rep. (in press) and references 
therein.
\bibitem  {d} S.H.Lee, C.Song and H.Yabu, Phys.Lett. \underline {B341}, 407 
(1995).
\bibitem  {xx} S.H.Lee, C.Song and C.M.Ko, Phys.Rev. \underline {C52}, R476 
(1995).
\bibitem   {e} G.Domokos and J.L.Goldman, Phys. Rev. \underline {D23},
 203 (1981).
\bibitem   {f} B.Sinha, Nucl. Phys. \underline {A459}, 717 (1986).
\bibitem  {g} P.V.Ruuskanen,  in Quark-gluon  plasma,  ed:  R.C.Hwa,  World 
    Scientific Publishing Company, Singapore, 1990.
\bibitem   {h} C.Gale and P.Lichard, Phys.Rev. \underline {D49}, 3338 (1994).
\bibitem   {i} J.D.Bjorken, Phys. Rev. \underline {D27}, 140 (1983).
\bibitem   {n} C.A.Dominguez, M.Loewe and J.S.Rozowsky, Phys.Lett. \underline 
{B335}, 506 (1994).
\bibitem   {o} D.Kharzeev and H.Satz, Phys.Lett. \underline {B340}, 167 
(1994).
\bibitem   {p} K.D.Born et.al., Phys.Rev.Lett. \underline {67}, 302 (1991).
\bibitem   {q} A.Gocksh, Phys.Rev.Lett. \underline {67}, 1701 (1991).
\end{thebibliography}
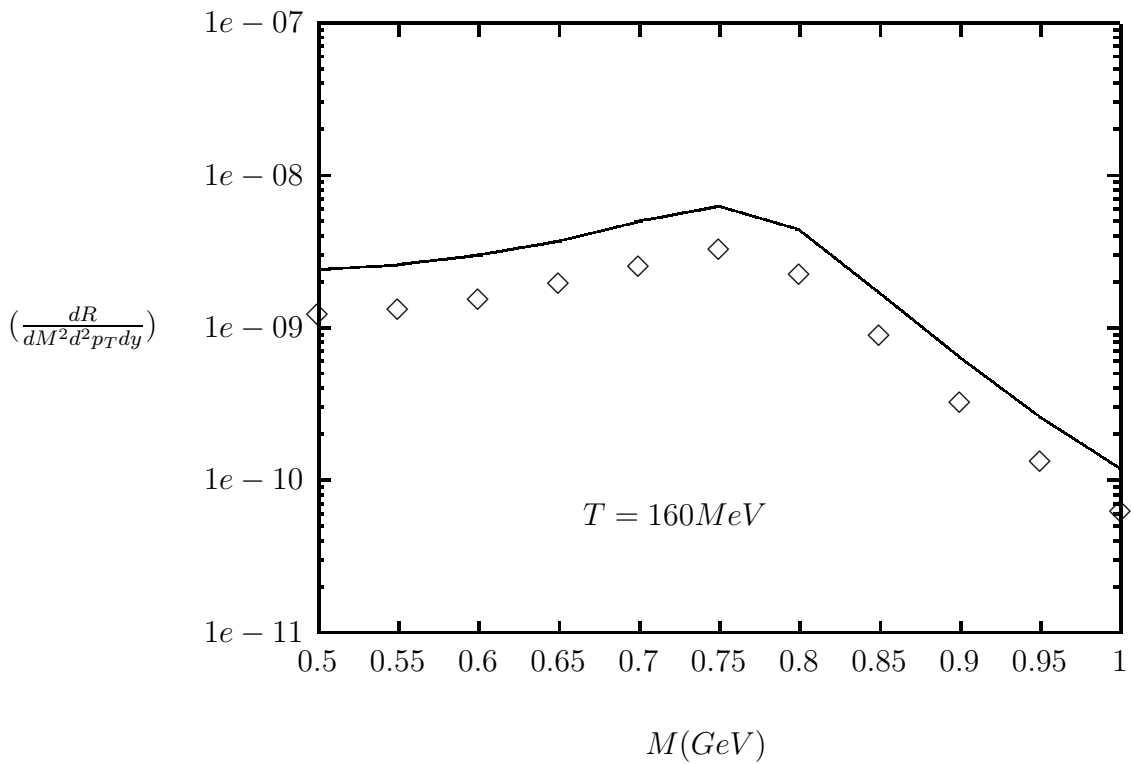
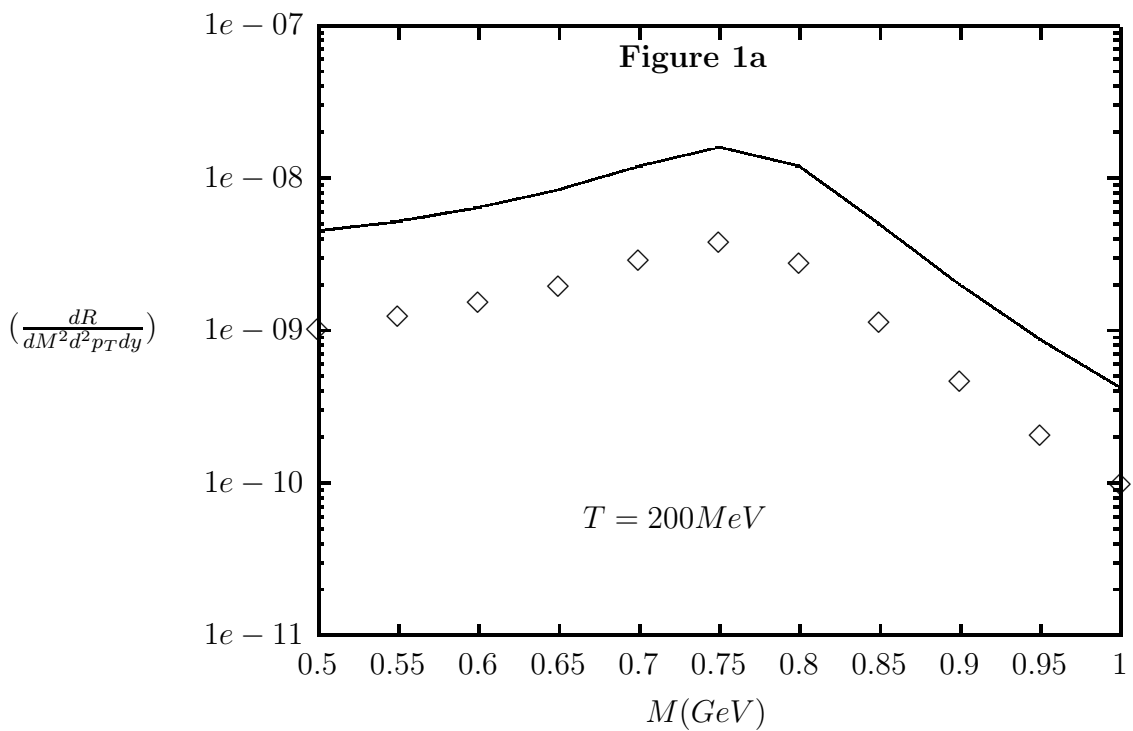
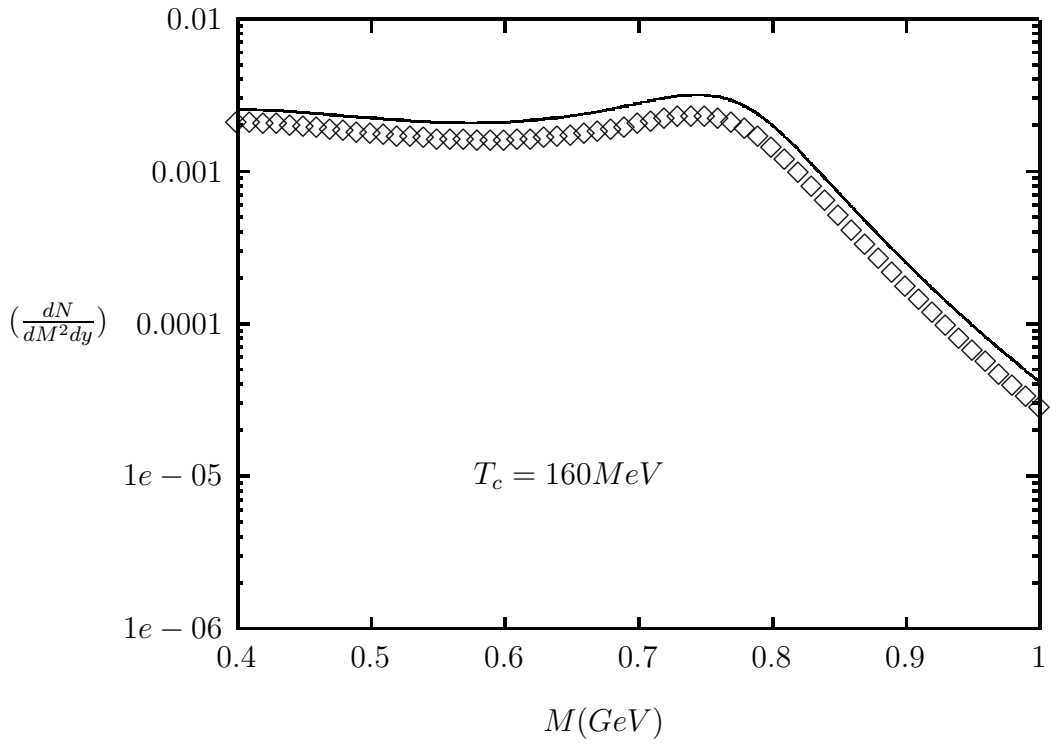
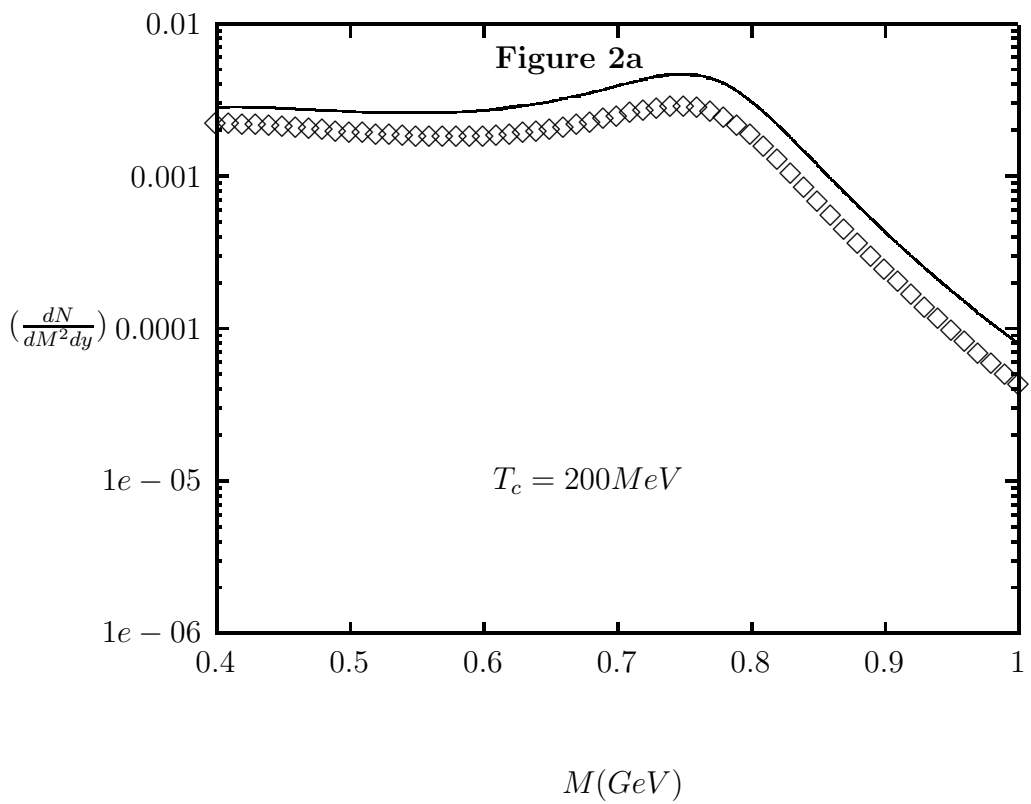
\end{document}